\documentclass[10pt,aps,prd,twocolumn,superscriptaddress,showpacs,preprintnumbers,amsmath,amssymb,floatfix,nofootinbib]{revtex4-1}

\usepackage{dcolumn}  
\usepackage[caption=false]{subfig}
\usepackage{color}
\usepackage{amssymb}
\usepackage{amsmath}
\usepackage{graphicx,bm}
\usepackage{verbatim}

\RequirePackage[colorlinks,citecolor=blue,urlcolor=blue,linkcolor=blue]{hyperref}
\RequirePackage{xspace}

\newcommand{\cms}{\ensuremath{{\rm CM}}\xspace}

\newcommand{\qb}{\ensuremath{Q_B}\xspace}
\newcommand{\ann}{NN\xspace}

\newcommand{\dn}{\ensuremath{D^0}\xspace}
\newcommand{\bn}{\ensuremath{B^0}\xspace}
\newcommand{\dbar}{\ensuremath{\overline{D}}\xspace}
\newcommand{\bbar}{\ensuremath{\overline{B}}\xspace}
\newcommand{\dnbar}{\ensuremath{\dbar{}^0}\xspace}
\newcommand{\bnbar}{\ensuremath{\bbar{}^0}\xspace}
\newcommand{\ks}{\ensuremath{K_S^0}\xspace}

\newcommand{\ups}{\ensuremath{\Upsilon(4S)}\xspace}
\newcommand{\mev}{\ensuremath{~{\rm MeV}}}
\newcommand{\gev}{\ensuremath{~{\rm GeV}}}
\newcommand{\mevc}{\ensuremath{~{\rm MeV/}c}}

\newcommand{\mevcsq}{\ensuremath{~{\rm MeV/}c^{2}}}
\newcommand{\gevcsq}{\ensuremath{~{\rm GeV/}c^{2}}}

\newcommand{\deld}{\ensuremath{\Delta\delta_D}\xspace}
\newcommand{\delf}{\ensuremath{\Delta\delta_f}\xspace}

\newcommand{\subsig}{\ensuremath{_{{\rm sig}}}\xspace}
\newcommand{\subtag}{\ensuremath{_{{\rm tag}}}\xspace}
\newcommand{\subbkg}{\ensuremath{_{{\rm bkg}}}\xspace}

\newcommand{\ndf}{\ensuremath{{\rm n.d.f.}}\xspace}

\newcommand{\dz}{\ensuremath{\Delta z}\xspace}
\newcommand{\dt}{\ensuremath{\Delta t}\xspace}
\newcommand{\mpsq}{\ensuremath{m_+^2}\xspace}
\newcommand{\mmsq}{\ensuremath{m_-^2}\xspace}

\newcommand{\cpconj}{\emph{CP}\xspace}

\newcommand{\pphi}    {\ensuremath{\varphi_1}\xspace}
\newcommand{\sindbeta}{\ensuremath{\sin{2\pphi}}\xspace}
\newcommand{\cosdbeta}{\ensuremath{\cos{2\pphi}}\xspace}
\newcommand{\grad}    {\ensuremath{^{\circ}}\xspace}

\newcommand{\ki}{\ensuremath{K_i}\xspace}
\newcommand{\kmi}{\ensuremath{K_{-i}}\xspace}
\newcommand{\kpmi}{\ensuremath{K_{\pm i}}\xspace}
\newcommand{\ci}{\ensuremath{C_i}\xspace}
\newcommand{\cmi}{\ensuremath{C_{-i}}\xspace}
\newcommand{\si}{\ensuremath{S_i}\xspace}
\newcommand{\smi}{\ensuremath{S_{-i}}\xspace}

\newcommand{\mcp}{\ensuremath{\mathcal{P}}\xspace}
\newcommand{\mcl}{\ensuremath{\mathcal{L}}\xspace}
\newcommand{\mck}{\ensuremath{\mathcal{K}}\xspace}
\newcommand{\mcd}{\ensuremath{\mathcal{D}}\xspace}
\newcommand{\mca}{\ensuremath{\mathcal{A}}\xspace}
\newcommand{\mcab}{\ensuremath{\overline{\mathcal{A}}}\xspace}

\newcommand{\ad}{\ensuremath{\mca_D}}
\newcommand{\af}{\ensuremath{\mca_f}\xspace}

\newcommand{\adbar}{\ensuremath{\mcab_D}}
\newcommand{\afbar}{\ensuremath{\mcab{}_f}\xspace}

\newcommand{\dvar}   {\ensuremath{\left(\mpsq,\mmsq\right)}\xspace}
\newcommand{\dvarinv}{\ensuremath{\left(\mmsq,\mpsq\right)}\xspace}

\newcommand{\pipi}{\ensuremath{\pi^+\pi^-}\xspace}
\newcommand{\kspp}{\ensuremath{\ks\pipi}\xspace}
\newcommand{\dkpp}{\ensuremath{D\to\kspp}\xspace}
\newcommand{\dnkpp}{\ensuremath{\dn   \to \kspp}\xspace}
\newcommand{\dbkpp}{\ensuremath{\dnbar\to \kspp}\xspace}
\newcommand{\ppp}{\ensuremath{\pipi\pi^0}\xspace}
\newcommand{\gaga}{\ensuremath{\gamma\gamma}\xspace}
\newcommand{\etap}{\ensuremath{\eta^{\prime}}\xspace}
\newcommand{\dst}{\ensuremath{D^{*0}}\xspace}
\newcommand{\dstbar}{\ensuremath{\dbar{}^{*0}}\xspace}
\newcommand{\bdh}{\ensuremath{\bn\to\dnbar h^0}\xspace}
\newcommand{\dstarh}{\ensuremath{D{}^{(*)0}h^0}\xspace}
\newcommand{\bdsth}{\ensuremath{\bn\to\dbar{}^{(*)0}h^0}\xspace}
\newcommand{\bdstarh}{\ensuremath{\bn\to\dstbar h^0}\xspace}
\newcommand{\bdpi}{\ensuremath{\bn\to\dnbar\pi^0}\xspace}
\newcommand{\dpi}{\ensuremath{\dnbar\pi^0}\xspace}
\newcommand{\bdeta}{\ensuremath{\bn\to\dnbar\eta}\xspace}
\newcommand{\deta}{\ensuremath{\dnbar\eta}\xspace}
\newcommand{\bdetagg}{\ensuremath{\bn\to\dnbar\eta_{\gaga}}\xspace}
\newcommand{\bdetap}{\ensuremath{\bn\to\dnbar\etap}\xspace}
\newcommand{\detap}{\ensuremath{\dnbar\etap}\xspace}
\newcommand{\bdetappp}{\ensuremath{\bn\to\dnbar\eta_{\ppp}}\xspace}
\newcommand{\bdomega}{\ensuremath{\bn\to\dnbar\omega}\xspace}
\newcommand{\domega}{\ensuremath{\dnbar\omega}\xspace}
\newcommand{\btodstpi}{\ensuremath{\bn\to\dstbar\pi^0}\xspace}
\newcommand{\dstpi}{\ensuremath{\dstbar\pi^0}\xspace}
\newcommand{\dsth}{\ensuremath{\dstbar h^0}\xspace}
\newcommand{\btodsteta}{\ensuremath{\bn\to\dstbar\eta}\xspace}
\newcommand{\dsteta}{\ensuremath{\dstbar\eta}\xspace}

\newcommand{\etapetapp}{\ensuremath{\etap\to[\gaga]_{\eta}\pipi}\xspace}
\newcommand{\dstdpi}{\ensuremath{\dst\to\dn\pi^0}\xspace}
\newcommand{\dstbdbpi}{\ensuremath{\dstbar\to\dnbar\pi^0}\xspace}
\newcommand{\dstdpip}{\ensuremath{D^{*+}\to \dn\pi^+}\xspace}
\newcommand{\ddbar}{\ensuremath{\dn\dnbar}\xspace}
\newcommand{\bbbar}{\ensuremath{B\bbar}\xspace}

\newcommand{\etagg}{\ensuremath{\eta\to\gaga}\xspace}

\newcommand{\hppp}{\ensuremath{h^0\to\ppp}\xspace}
\newcommand{\etappp}{\ensuremath{\eta\to\ppp}\xspace}

\newcommand{\omegappp}{\ensuremath{\omega\to\ppp}\xspace}

\newcommand{\btau}{\ensuremath{\tau_B}\xspace}
\newcommand{\bexp}{\ensuremath{e^{-\frac{\left|\Delta t\right|}{\btau}}}\xspace}
\newcommand{\lamfmsq}{\ensuremath{\left|\lambda_f\right|^2}\xspace}
\newcommand{\lamf}{\ensuremath{\lambda_f}\xspace}

\newcommand{\bptodpi}{\ensuremath{B^{+}\to\dnbar\pi^+}\xspace}
\newcommand{\bptodk}{\ensuremath{B^{+}\to\dnbar K^+}\xspace}

\newcommand{\stat}{\ensuremath{\left({\rm stat.}\right)}\xspace}
\newcommand{\syst}{\ensuremath{\left({\rm syst.}\right)}\xspace}

\newcommand{\de}{\ensuremath{\Delta E}\xspace}
\newcommand{\mbc}{\ensuremath{M_{\rm bc}}\xspace}
\newcommand{\dembc}{\de--\mbc}
\newcommand{\dmb}{\ensuremath{\Delta m_B}\xspace}
\newcommand{\dmdt}{\ensuremath{\left(\dmb\dt\right)}\xspace}

\newcommand{\btoccs}{\ensuremath{b\to c\overline{c}s}\xspace}

\newcommand{\btocud}{\ensuremath{b\to c\overline{u}d}\xspace}

\newcommand{\bbartocud}{\ensuremath{\overline{b}\to\overline{c}u\overline{d}}\xspace}
\newcommand{\bbartoucd}{\ensuremath{\overline{b}\to \overline{u}c\overline{d}}\xspace}

{}
\begin{document}

\title{\boldmath Measurement of the CKM angle \pphi in \bdsth, \dbkpp decays with time-dependent binned Dalitz plot analysis}

\noaffiliation
\affiliation{University of the Basque Country UPV/EHU, 48080 Bilbao}
\affiliation{Beihang University, Beijing 100191}
\affiliation{Budker Institute of Nuclear Physics SB RAS, Novosibirsk 630090}
\affiliation{Faculty of Mathematics and Physics, Charles University, 121 16 Prague}
\affiliation{Chonnam National University, Kwangju 660-701}
\affiliation{University of Cincinnati, Cincinnati, Ohio 45221}
\affiliation{Deutsches Elektronen--Synchrotron, 22607 Hamburg}
\affiliation{University of Florida, Gainesville, Florida 32611}
\affiliation{Justus-Liebig-Universit\"at Gie\ss{}en, 35392 Gie\ss{}en}
\affiliation{SOKENDAI (The Graduate University for Advanced Studies), Hayama 240-0193}
\affiliation{Hanyang University, Seoul 133-791}
\affiliation{University of Hawaii, Honolulu, Hawaii 96822}
\affiliation{High Energy Accelerator Research Organization (KEK), Tsukuba 305-0801}
\affiliation{J-PARC Branch, KEK Theory Center, High Energy Accelerator Research Organization (KEK), Tsukuba 305-0801}
\affiliation{IKERBASQUE, Basque Foundation for Science, 48013 Bilbao}
\affiliation{Indian Institute of Science Education and Research Mohali, SAS Nagar, 140306}
\affiliation{Indian Institute of Technology Bhubaneswar, Satya Nagar 751007}
\affiliation{Indian Institute of Technology Guwahati, Assam 781039}
\affiliation{Indian Institute of Technology Madras, Chennai 600036}
\affiliation{Indiana University, Bloomington, Indiana 47408}
\affiliation{Institute of High Energy Physics, Chinese Academy of Sciences, Beijing 100049}
\affiliation{Institute of High Energy Physics, Vienna 1050}
\affiliation{Institute for High Energy Physics, Protvino 142281}
\affiliation{INFN - Sezione di Torino, 10125 Torino}
\affiliation{J. Stefan Institute, 1000 Ljubljana}
\affiliation{Kanagawa University, Yokohama 221-8686}
\affiliation{Institut f\"ur Experimentelle Kernphysik, Karlsruher Institut f\"ur Technologie, 76131 Karlsruhe}
\affiliation{Kennesaw State University, Kennesaw, Georgia 30144}
\affiliation{King Abdulaziz City for Science and Technology, Riyadh 11442}
\affiliation{Korea Institute of Science and Technology Information, Daejeon 305-806}
\affiliation{Korea University, Seoul 136-713}
\affiliation{Kyungpook National University, Daegu 702-701}
\affiliation{\'Ecole Polytechnique F\'ed\'erale de Lausanne (EPFL), Lausanne 1015}
\affiliation{P.N. Lebedev Physical Institute of the Russian Academy of Sciences, Moscow 119991}
\affiliation{Faculty of Mathematics and Physics, University of Ljubljana, 1000 Ljubljana}
\affiliation{Ludwig Maximilians University, 80539 Munich}
\affiliation{Luther College, Decorah, Iowa 52101}
\affiliation{University of Maribor, 2000 Maribor}
\affiliation{Max-Planck-Institut f\"ur Physik, 80805 M\"unchen}
\affiliation{School of Physics, University of Melbourne, Victoria 3010}
\affiliation{University of Miyazaki, Miyazaki 889-2192}
\affiliation{Moscow Physical Engineering Institute, Moscow 115409}
\affiliation{Moscow Institute of Physics and Technology, Moscow Region 141700}
\affiliation{Graduate School of Science, Nagoya University, Nagoya 464-8602}
\affiliation{Nara Women's University, Nara 630-8506}
\affiliation{National Central University, Chung-li 32054}
\affiliation{National United University, Miao Li 36003}
\affiliation{Department of Physics, National Taiwan University, Taipei 10617}
\affiliation{H. Niewodniczanski Institute of Nuclear Physics, Krakow 31-342}
\affiliation{Nippon Dental University, Niigata 951-8580}
\affiliation{Niigata University, Niigata 950-2181}
\affiliation{Novosibirsk State University, Novosibirsk 630090}
\affiliation{Pacific Northwest National Laboratory, Richland, Washington 99352}
\affiliation{University of Pittsburgh, Pittsburgh, Pennsylvania 15260}
\affiliation{Punjab Agricultural University, Ludhiana 141004}
\affiliation{Theoretical Research Division, Nishina Center, RIKEN, Saitama 351-0198}
\affiliation{University of Science and Technology of China, Hefei 230026}
\affiliation{Showa Pharmaceutical University, Tokyo 194-8543}
\affiliation{Soongsil University, Seoul 156-743}
\affiliation{Sungkyunkwan University, Suwon 440-746}
\affiliation{School of Physics, University of Sydney, New South Wales 2006}
\affiliation{Department of Physics, Faculty of Science, University of Tabuk, Tabuk 71451}
\affiliation{Tata Institute of Fundamental Research, Mumbai 400005}
\affiliation{Excellence Cluster Universe, Technische Universit\"at M\"unchen, 85748 Garching}
\affiliation{Department of Physics, Technische Universit\"at M\"unchen, 85748 Garching}
\affiliation{Toho University, Funabashi 274-8510}
\affiliation{Department of Physics, Tohoku University, Sendai 980-8578}
\affiliation{Earthquake Research Institute, University of Tokyo, Tokyo 113-0032}
\affiliation{Department of Physics, University of Tokyo, Tokyo 113-0033}
\affiliation{Tokyo Institute of Technology, Tokyo 152-8550}
\affiliation{Tokyo Metropolitan University, Tokyo 192-0397}
\affiliation{Virginia Polytechnic Institute and State University, Blacksburg, Virginia 24061}
\affiliation{Wayne State University, Detroit, Michigan 48202}
\affiliation{Yamagata University, Yamagata 990-8560}
\affiliation{Yonsei University, Seoul 120-749}
  \author{V.~Vorobyev}\affiliation{Budker Institute of Nuclear Physics SB RAS, Novosibirsk 630090}\affiliation{Novosibirsk State University, Novosibirsk 630090} 
  \author{I.~Adachi}\affiliation{High Energy Accelerator Research Organization (KEK), Tsukuba 305-0801}\affiliation{SOKENDAI (The Graduate University for Advanced Studies), Hayama 240-0193} 
  \author{H.~Aihara}\affiliation{Department of Physics, University of Tokyo, Tokyo 113-0033} 
  \author{D.~M.~Asner}\affiliation{Pacific Northwest National Laboratory, Richland, Washington 99352} 
  \author{T.~Aushev}\affiliation{Moscow Institute of Physics and Technology, Moscow Region 141700} 
  \author{R.~Ayad}\affiliation{Department of Physics, Faculty of Science, University of Tabuk, Tabuk 71451} 
  \author{I.~Badhrees}\affiliation{Department of Physics, Faculty of Science, University of Tabuk, Tabuk 71451}\affiliation{King Abdulaziz City for Science and Technology, Riyadh 11442} 
  \author{S.~Bahinipati}\affiliation{Indian Institute of Technology Bhubaneswar, Satya Nagar 751007} 
  \author{A.~M.~Bakich}\affiliation{School of Physics, University of Sydney, New South Wales 2006} 
  \author{P.~Behera}\affiliation{Indian Institute of Technology Madras, Chennai 600036} 
% ...
\collaboration{The Belle Collaboration}

\begin{abstract}
We report a measurement of the \cpconj violation parameter \pphi obtained in a 
time-dependent analysis of \bdsth decays followed by \dbkpp decay. A model-independent 
measurement is performed using the binned Dalitz plot technique. The measured value is 
$\pphi = 11.7\grad\pm7.8\grad\stat\pm 2.1\grad\syst$.  Treating \sindbeta and 
\cosdbeta as independent parameters, we obtain $\sindbeta = 0.43\pm 0.27\stat\pm 0.08\syst$ 
and $\cosdbeta = 1.06\pm 0.33\stat^{+0.21}_{-0.15}\syst$.  The results are obtained with a 
full data sample of $772 \times 10^6 \bbbar$ pairs collected near the \ups resonance with 
the Belle detector at the KEKB asymmetric-energy $e^+e^-$ collider.
\end{abstract}

\pacs{11.30.Er, 13.25.Gv, 13.25.Hw}

\maketitle
\section{Introduction}
The study of \cpconj symmetry provides valuable insight into the structure and dynamics 
of matter from the subatomic to the cosmic scale.  \cpconj violation is a necessary 
ingredient for baryogenesis and explaining the state of matter in the observable 
Universe~\cite{sakharov}.  The Standard Model (SM) of particle physics accounts for \cpconj 
violation using the mechanism proposed by Kobayashi and Maskawa (KM)~\cite{KM}.  A unitary 
matrix of quark flavor mixing, referred to as the Cabibbo-Kobayashi-Maskawa 
(CKM)~\cite{cabibbo,KM} matrix, encodes this mechanism. The CKM matrix makes charged weak 
currents non-invariant under \cpconj transformation. The SM does not predict the values 
of the elements of the CKM matrix, but theoretical predictions estimate that the amount 
of \cpconj violation introduced by the SM is too feeble to explain the baryon asymmetry 
of the Universe~\cite{baryogenesis}. Thus, it is important to test the KM mechanism and 
search for new sources of \cpconj violation.

\begin{figure}[htb]
\subfloat[]{\label{fig:bcud}\includegraphics[width=0.23\textwidth]{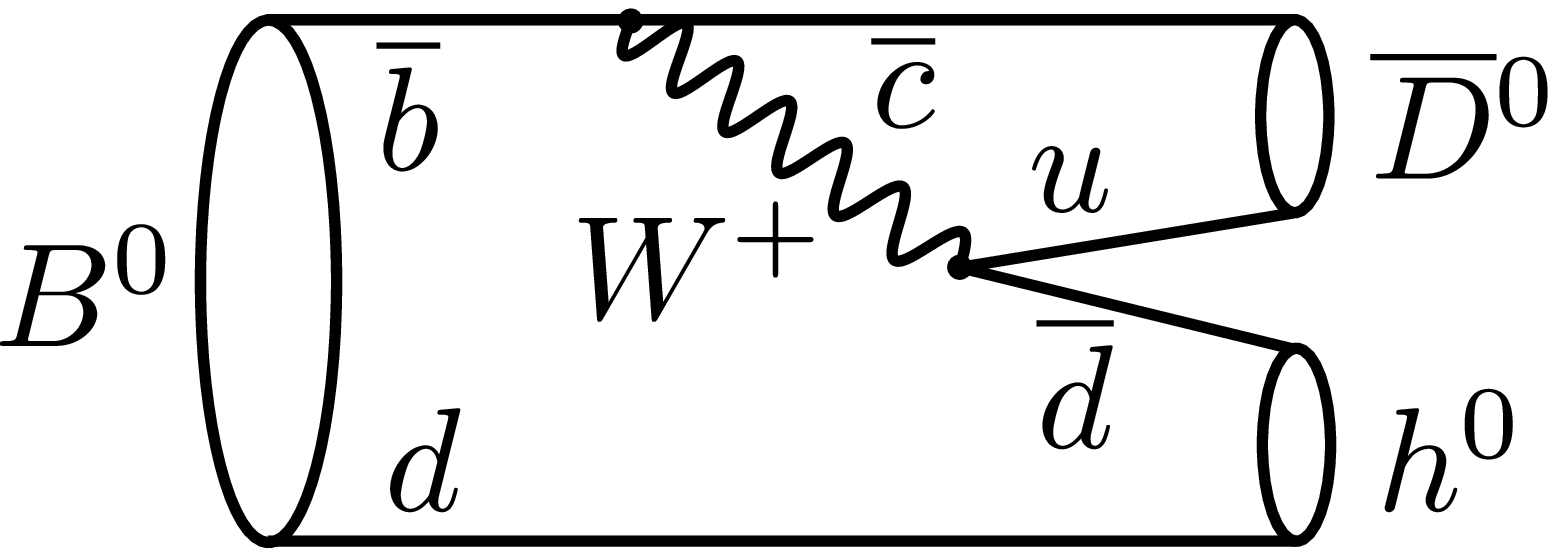}}\hfill
\subfloat[]{\label{fig:bucd}\includegraphics[width=0.23\textwidth]{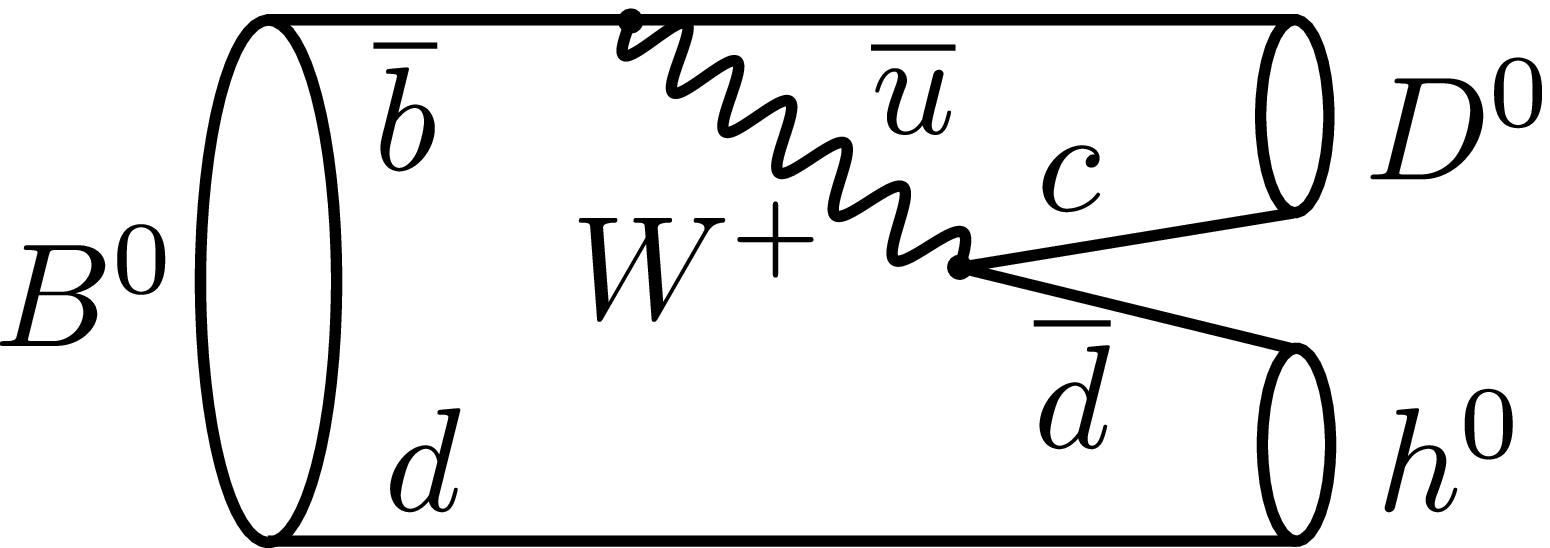}}
\caption{\bbartocud transition~(a) leading to \bdh decay and \bbartoucd transition~(b) 
leading to $\bn\to\dn h^0$ decay.}
\label{fig:bdh}
\end{figure}

Unitarity of the CKM matrix implies several relations among its elements that can be 
represented as triangles in the complex plane.  In particular, the relation formed by 
the elements of the first and the third columns, referred to as the Unitarity 
Triangle~(UT)~\cite{ut}, is the most accessible for experimental tests.  

The \cpconj violation parameter $\pphi=\arg(-\,V_{cd}V_{cb}^*/V_{td}V_{tb}^*)$, where 
$V_{ij}$ is an element of the CKM matrix, is one of the angles of the UT\footnote{Another 
naming convention, $\beta$ ($\equiv\pphi$), is also used in the literature.}.  The value 
of \sindbeta has been measured precisely in \btoccs transitions by Belle, BaBar and 
LHCb~\cite{sinbeta}.  Two discrete ambiguities remain with the known value of 
\sindbeta: $\pphi\to \pphi+\pi$ and $\pphi\to \pi/2-\pphi$.  Currently, no theoretical 
approach is available to resolve the former ambiguity, but the latter can be resolved by 
measuring \cosdbeta.  Existing measurements of \cosdbeta in \btocud~\cite{BaBarbdh0,belle_bdh0} 
and \btoccs~\cite{cosbeta_jphikst,cosbeta_dstdstks} transitions are much less precise and, 
in most cases, model-dependent.  

Here, we present a model-independent measurement of the angle \pphi in \btocud 
transitions (Fig.~\ref{fig:bcud}) governing \bdsth decays with subsequent \dbkpp 
decay\footnote{Throughout this paper, the inclusion of the charge-conjugate decay mode 
is implied unless otherwise stated.}, where $h^0$ is a light unflavored meson.  
This measurement is based on a data sample twice as large as that used in the previous 
\pphi measurement using \bdsth decays at Belle~\cite{belle_bdh0}.  The technique of a binned 
Dalitz plot analysis is applied to the \pphi measurement for the first time.

\subsection{Formalism}
This section describes the technique to measure the angle \pphi at an asymmetric-energy 
$e^+e^-$ collider operating at center-of-mass (\cms) energy near the \ups 
resonance~\cite{Sanda,BPhys}.  When a pair of neutral $B$ mesons is produced, they  
oscillate coherently until one decays.  Therefore, at the moment of a flavor-specific 
decay of one of the $B$ mesons (in the \ups rest frame), the flavor of the other $B$ meson 
is fixed.  The former $B$ meson is referred to as the {\it tagging} $B$ meson and the 
latter as the {\it signal} $B$ meson.  The tagging and signal $B$ mesons decay 
at proper times $t\subtag$ and $t\subsig$, respectively.  

The longitudinal distance \dz along the beam axis between the decay vertices of the
signal and tagging $B$ mesons in the lab frame is measured. Since the $B$ mesons 
are produced almost at rest in the \cms frame, their momentum can be neglected 
and the approximation $\dt\approx\dz/(c\beta\gamma)$ can be used, where 
$\dt=t\subsig-t\subtag$ and $\beta$ and $\gamma$ are the Lorentz factors of 
the \ups parent.

If the amplitudes $\mca\left(\bn\to f\right)\equiv\af$ and $\mca\left(\bnbar\to f\right)
\equiv\afbar$ are non-zero for some final state~$f$, then the distribution of the decay 
time difference, attributed to the interference of the processes $\bn\to~f$ and 
$\bn~\to~\bnbar~\to~f$, is~\cite{Sanda}
\begin{equation}\label{eq:mod_dep_pdf}
\begin{split}
 \mcp(\dt) &= h_1\bexp\left[1 +\frac{1-\lamfmsq}{1+\lamfmsq}\cos{\dmdt}\right.\\ 
               &\left.-\frac{2\,{\rm Im}\,\lamf}{1+\lamfmsq}\sin{\dmdt}\right],\quad
               \lamf=\frac{q}{p}\frac{\afbar}{\af},
\end{split}
\end{equation}
where 
$p$ and $q$ are 
the coefficients relating the mass and flavor $B$-meson eigenstates to each other, 
\btau is the neutral $B$ meson lifetime (assumed to be the same for both mass eigenstates), 
\dmb is the mass difference between the mass eigenstates, and $h_1$ is a normalizing constant.  
In the following, we assume the absence of \cpconj violation in mixing and a null 
\cpconj-violating weak phase in the $B$ meson decay amplitudes:
\begin{equation}\label{eq:phi1}
 \frac{q}{p} = e^{-i2\pphi},\quad \arg\left(\frac{\afbar}{\af}\right) = \delf,
\end{equation}
so that
\begin{equation}\label{eq:imlamf}
 {\rm Im}\,\lamf = \left|\frac{\afbar}{\af}\right|\sin{\left(\delf-2\pphi\right)};
\end{equation}
here, \delf is the difference in strong phases, which does not change sign under 
a \cpconj transformation.  Consideration of the \cpconj-conjugated process, in which 
the \cpconj-violating phase \pphi is replaced by $-\pphi$, allows one to distinguish 
between the weak ($2\pphi$) and strong (\delf) phases.

For \bdsth decays, the amplitudes \af and \afbar can be expressed as
\begin{equation}\label{eq:bdh_amp}
 \af = \alpha_B\adbar,\quad \afbar = \alpha_B\xi_{h^0}\left(-1\right)^L\ad,
\end{equation}
where $\xi_{h^0}$ is the \cpconj eigenvalue of the $h^0$ meson, $L$ is 
the relative angular momentum in the \dstarh system, \ad\ (\adbar) is the \dn (\dnbar) 
decay amplitude into the final state $f_D$, and $\alpha_B$ is a complex coefficient.  
The charm mixing and possible \cpconj violation in the $D$ meson decays are neglected 
in Eq.~(\ref{eq:bdh_amp}). With the existing $B$-factories statistics, the $\bn\to\dn h^0$ 
decay amplitude (Fig.~\ref{fig:bucd}) can be neglected with respect to the \bdh decay 
amplitude (Fig.~\ref{fig:bcud}) because it is suppressed by $\left|V_{ub}V_{cd}/V_{cb}
V_{ud}\right|\approx 0.02$.  

If the state $f_D$ is a \cpconj eigenstate, then the entire state $f$ is \cpconj eigenstate 
(except for the \dsth state with a vector $h^0$ meson) as well and the phase \delf equals 
$0$ or~$\pi$. This exposes a sensitivity to \sindbeta but not \cosdbeta and provides the 
best way to measure \sindbeta in \btocud transitions~\cite{markus_dcp}.

The three-body state $f_D=\kspp$ is not a \cpconj eigenstate, so the phase \delf is not 
limited to the values $0$ and $\pi$.  As a consequence, this state provides sensitivity 
to both \sindbeta and \cosdbeta.  The amplitude of \dnkpp decay can be expressed as a 
function \ad\dvar of two Dalitz-distribution variables~\cite{dalitz}, where $m_{\pm} = 
m\left(\ks\pi^{\pm}\right)$ are the invariant masses. The amplitude $\adbar$ of \dbkpp 
decay can be obtained by transposing the Dalitz variables: $\adbar\dvar\equiv\ad\dvarinv$.  
Therefore, the phase difference \delf is a function of the Dalitz variables:
\begin{equation}
 \begin{split}
  \delf\dvar &= \arg\left(\xi_{h^0}(-1)^L\right)-\deld\dvar,\\
  \deld\dvar &= \arg\left(\frac{\ad\dvarinv}{\ad\dvar}\right).
 \end{split}
\end{equation}

For the $f_D=\kspp$ final state, the strong phase \deld cannot be measured at each point 
in the phase space: additional information is necessary. An approach based on an isobar 
model of the $D$ meson decay amplitude was proposed in Ref.~\cite{BGK} and used in the 
measurement of the CKM angle \pphi performed by BaBar~\cite{BaBarbdh0} and Belle~\cite{belle_bdh0}.  
Alternatively, we use here a method that is independent of the decay model, as described below.

\subsection{Time-dependent binned Dalitz plot analysis}
Our measurement is based on the binned Dalitz distribution approach.  This idea was proposed 
in Ref.~\cite{GGSZ} to measure the CKM angle $\varphi_3$ and further developed for several 
applications in Refs.~\cite{BP_phi3_model,BPV,hr}.  We extend this approach to measure the 
angle \pphi in the time-dependent analysis of \bdsth, \dbkpp decays.  The Dalitz plot is 
divided into $16$ bins ($2\mathcal{N}$ in the general case) symmetrically with respect 
to $\mpsq\leftrightarrow\mmsq$ exchange.  The bin index $i$ lies between $-8$ and $8$, 
excluding~$0$; $\mpsq\leftrightarrow\mmsq$ exchange corresponds to the sign inversion $i\to -i$.

Several parameters related to a Dalitz plot bin on the Dalitz plane \mcd 
are introduced.  These are the probability for the \dnbar meson to decay into the 
phase space region $\mcd_i$ of the Dalitz plot bin $i$
\begin{equation}\label{eq:k}
 \ki = \int\limits_{\mcd_i}\left|\ad\dvarinv\right|^2 d\mpsq d\mmsq,
\end{equation}
(normalized by $\sum_{i=-8}^{8}{\ki} = 1$) and the weighted averages of the sine and cosine of 
the phase difference between \dnbar and \dn decay amplitudes \deld\dvar over the $i$-th Dalitz 
plot bin:
  \begin{equation}\label{eq:cs}
  \begin{split}
   &\ci=\frac{\int\limits_{\mcd_i}
            \left|\ad\right|\left|\adbar\right|
            \cos\deld\,d\mpsq d\mmsq
            }{\sqrt{
            \ki\kmi
            }},\\
  &\si=\frac{\int\limits_{\mcd_i}
            \left|\ad\right|\left|\adbar\right|
            \sin\deld\,d\mpsq d\mmsq
            }{\sqrt{
            \ki\kmi
            }}\,.
 \end{split}
\end{equation}
The binning method yields the relations $\ci=\cmi$ and $\si=-\smi$.  Eq.~(\ref{eq:mod_dep_pdf}) 
can be expressed in the form appropriate for a time-dependent binned analysis:
\begin{widetext}
\begin{equation}\label{eq:master-formula}
 \begin{split}
  \mcp_i\left(\dt,\pphi\right) &= h_2\bexp\left[ 1 + q_B\frac{\ki-\kmi}{\ki+\kmi}\cos\dmdt\right.\\
  &\left.+2q_B\xi_{h^0}(-1)^L\frac{\sqrt{\ki\kmi}}{\ki+\kmi}\sin\dmdt\left(\si\cosdbeta+\ci\sindbeta\right)\right],
 \end{split}
 \end{equation} 
\end{widetext}
where $q_B = -1$ ($+1$) corresponds to a signal \bn (\bnbar) meson and $h_2$ is a normalizing constant.  

The knowledge of the signal-event distribution over the Dalitz plot bins for both $B$ meson flavors 
is necessary for the fit that extracts the \cpconj violation parameters.  The expected fraction 
$n_{i,q_B}$ of signal events for the $i$-th Dalitz plot bin and signal $B$ flavor $q_B$ is
\begin{equation}\label{eq:nsig-prediction-with-tag}
  n_{i,q_B} =\frac{\ki + \kmi}{2} + \frac{q_B}{1+(\btau\dmb)^2}\cdot\frac{\ki - \kmi}{2}.
\end{equation}
This formula is obtained by integrating Eq.~(\ref{eq:master-formula}) over \dt.

In principle, each pair $(i,-i)$ of bins provides enough 
information to measure the \cpconj violation parameters if the values of parameters 
\kpmi, \ci, and \si are known and do not equal zero.
\begin{figure*}[htb]
\subfloat[]{\label{fig:dp_mc}%
  \includegraphics[width=0.45\textwidth]{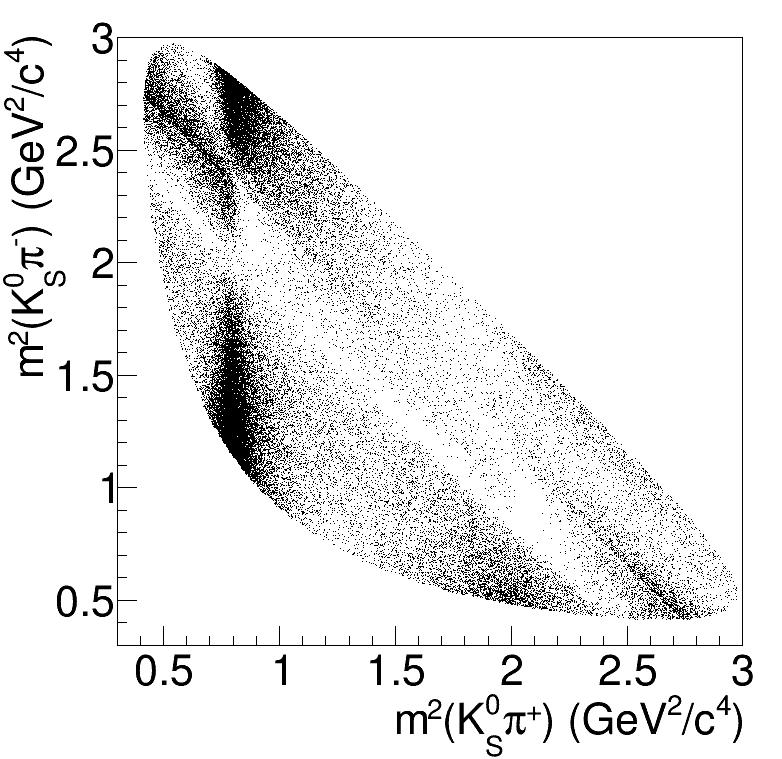}%
}\hfill
\subfloat[]{\label{fig:binned_dp}%
  \includegraphics[width=0.45\textwidth]{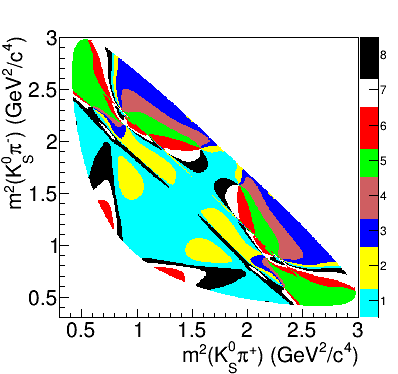}%
}
\caption{Dalitz plot distribution (a) and equal-phase binning (b) obtained with 
the amplitude model of \dbkpp decay from~Ref.~\cite{Belle_model}.}
\label{fig:Belle_model}
\end{figure*}

For a given binning of the Dalitz plot, the parameters \ki can be measured with a set of 
flavor-tagged neutral $D$ mesons such as \dstdpip or \bptodpi decays, by measuring signal 
yield in each Dalitz plot bin.  The measurement of the phase parameters \ci and \si is 
more complicated and can be done with coherent decays of \ddbar pairs~\cite{CLEO_phasees}.

Measurement of the \cpconj violation parameters is possible for an arbitrary binning 
of the Dalitz plot, but usage of the realistic decay amplitude model allows one to optimize 
the binning to approach the maximal statistical sensitivity.  In particular, the 
equal-phase binning method \cite{BP_phi3_model} suggests the following rule for $i>0$ 
and $\mpsq<\mmsq$:
\begin{equation}\label{eq:bin_condition}
\frac{\pi(i-3/2)}{4}<\deld\left(m_+^2,m_-^2\right)<\frac{\pi(i-1/2)}{4}.
\end{equation}
This binning and the \dnkpp decay amplitude model reported in~Ref.~\cite{Belle_model} 
(see Fig.~\ref{fig:Belle_model}) are employed in the analysis presented here.  
The analysis uses the values of \ki extracted from the \bptodpi sample, as described 
in Section~\ref{sec:bdpip}, and the values of \ci and \si parameters measured by 
CLEO-c~\cite{CLEO_phasees}, as listed in Table~\ref{tab:CLEO_measurements}.

Model-inspired binning of the Dalitz plot does not lead to a bias in the measured parameters, 
because of the excellent invariant mass resolution of the detector. Therefore, an alternative 
binning derived from a model that parameterized the data poorly would only reduce the statistical 
sensitivity of the measurement.

\begin{table}[htb]
 \caption{The values of the parameters \ci and \si measured by CLEO-c~\cite{CLEO_phasees} 
 for equal-phase Dalitz-plot binning according to the \dbkpp decay model obtained in~Ref.~\cite{Belle_model}.}
 \label{tab:CLEO_measurements}
 \begin{tabular}
  { @{\hspace{0.2cm}}c@{\hspace{0.2cm}} @{\hspace{0.2cm}}r@{\hspace{0.2cm}} @{\hspace{0.2cm}}r@{\hspace{0.2cm}}} \hline\hline
  Bin &  \multicolumn{1}{c}{\ci} & \multicolumn{1}{c}{\si}\\ \hline
  $1$ & $ 0.710\pm 0.034\pm 0.038$ & $-0.013\pm 0.097\pm 0.031$ \\
  $2$ & $ 0.481\pm 0.080\pm 0.070$ & $-0.147\pm 0.177\pm 0.107$ \\
  $3$ & $ 0.008\pm 0.080\pm 0.087$ & $ 0.938\pm 0.120\pm 0.047$ \\
  $4$ & $-0.757\pm 0.099\pm 0.065$ & $ 0.386\pm 0.208\pm 0.067$ \\
  $5$ & $-0.884\pm 0.056\pm 0.054$ & $-0.162\pm 0.130\pm 0.041$ \\
  $6$ & $-0.462\pm 0.100\pm 0.082$ & $-0.616\pm 0.188\pm 0.052$ \\
  $7$ & $ 0.106\pm 0.105\pm 0.100$ & $-1.063\pm 0.174\pm 0.066$ \\
  $8$ & $ 0.365\pm 0.071\pm 0.078$ & $-0.179\pm 0.166\pm 0.048$ \\ \hline
  \hline
 \end{tabular}
\end{table}

\section{Belle detector}
This measurement is based on a data sample that contains $772 \times 10^6~\bbbar$ pairs, 
collected with the Belle detector at the KEKB asymmetric-energy $e^+e^-$ ($3.5$ on $8$~GeV) 
collider~\cite{KEKB} operated near the \ups resonance.

The Belle detector is a large-solid-angle magnetic
spectrometer that consists of a silicon vertex detector (SVD)
featuring the double-sided silicon strip devices,
a $50$-layer central drift chamber (CDC),
an array of aerogel threshold Cherenkov counters, 
a barrel-like arrangement of time-of-flight
scintillation counters, 
and an electromagnetic calorimeter
comprised of CsI(Tl) crystals 
located inside a super-conducting solenoid coil that provides a $1.5$~T
magnetic field.  An iron flux-return located outside of
the coil is instrumented to detect $K_L^0$ mesons and to identify muons.  
The detector is described in detail elsewhere~\cite{Belle}.  
Two inner detector configurations were used. A $2.0$~cm radius beampipe
and a $3$-layer silicon vertex detector was used for the first sample
of $152 \times 10^6~\bbbar$ pairs, while a $1.5$~cm radius beampipe, a $4$-layer
silicon vertex detector and a small-cell inner drift chamber were used to record  
the remaining $620 \times 10^6~\bbbar$ pairs~\cite{svd2}.  

\section{Event selection}
Six \bdsth decay modes, \dpi, \deta, \detap, \domega, \dstpi, and \dsteta, 
with subsequent decays \dbkpp, \etagg or \ppp, \omegappp, \etapetapp, and 
\dstbdbpi, are used in this analysis.  Only \etagg is considered for the 
\detap and \dsteta modes.  Charged $B$-meson decay \bptodpi followed by 
\dbkpp is used to measure the parameters~\ki.

The charged pion candidates are selected from the reconstructed tracks and are required 
to have both $z$ and $r\varphi$ hits in at least one layer and at least one additional 
layer with a $z$ hit.  The impact parameters of the tracks with respect to the beam 
interaction point in the longitudinal and transverse projections are required to satisfy 
$\left|dz\right|<5$~cm and $dr<2$~cm, respectively.  The transverse momentum $p_t$ is 
required to be greater than $50\mevc$ ($100\mevc$) for pions produced in \dnkpp (\hppp) 
decay.  These requirements are not applied for the pions daughters of~\ks candidates.

The~$\ks\to\pipi$ candidates are reconstructed from two oppositely charged 
tracks using two artificial neural networks~(\ann).  The first \ann is trained 
to suppress the combinatorial background and fake tracks: it uses 
the track impact parameters with respect to the beam interaction point,
the azimuthal angle between the \ks momentum and the decay-vertex vectors,
the distance between the tracks,
the \ks flight length in the $x$-$y$~plane,
the \ks momentum, the distance between the beam interaction point and the tracks, 
the angle between the \ks and a pion flight directions, 
the presence of the SVD hits and number of CDC hits on the tracks.  
The second \ann is trained to suppress the background from $\Lambda\to p\pi^-$ 
decays: it uses the reconstructed mass with the lambda hypothesis, 
the absolute values of the track momenta, the track-momenta polar angles and 
the particle identification parameter distinguishing pions from protons. Further details 
of the procedure are described in Ref.~\cite{nisks}.  The invariant mass of the selected 
candidates is required to be between $488.5$ and $506.5\mevcsq$.  This mass interval, as 
well as any other mass interval used in the analysis (unless explicitly stated otherwise), 
correspond to $\pm3$ standard deviations from the nominal value.
 
The $\pi^0$ candidates are formed from photon pairs with an invariant mass between 
$115.7$ and $153.7\mevcsq$.  The photon energy is required to be greater than $40\mev$.  
The energy of the $\pi^0$ candidate from \hppp ($h^0=\eta$ and $\omega$) decay must to 
be greater than $200\mev$.
 
The \etagg candidates are formed from photon pairs with an invariant mass between $530.0$ 
and $573.7\mevcsq$.  The photon energy is required to be greater than $80\mev$.
 
The \hppp candidates, where $h^0~=~\eta$ or $\omega$, are formed from a $\pi^0$ 
candidate and two oppositely charged tracks with invariant mass between $537.6$ 
and $557.4\mevcsq$ for $\eta$ and between $760.4$ and $803.9\mevcsq$ for $\omega$.  
For the $\omega$ candidates, the absolute value of the cosine of the helicity angle 
$\theta_{\rm hel}$ (the angle between the \bn flight direction and the normal to the 
$\omega$ decay plane in the $\omega$ rest frame) is required to be greater than~$0.2$.
 
The $\etap\to\eta\pi^+\pi^-$ candidates are formed from a \etagg candidate and two 
oppositely charged tracks, both treated as pions.  The invariant mass difference 
$\Delta m_{\eta}~\equiv~m(\etap)-m(\eta)$ is required to lie between $401.7$ and~$417.7\mevcsq$.

The \dnkpp candidates are formed from a \ks candidate and two oppositely charged tracks, 
both treated as pions, with an invariant mass between $1.8516$ and~$1.8783\gevcsq$.
 
The \dstdpi candidates are formed from a \dn candidate and a neutral pion candidate.  The 
invariant mass difference $\Delta m_{D} \equiv m(\dst)-m(\dn)$ must lie between $140.2$ and~$144.2\mevcsq$.

The selection of \bn and $B^{\pm}$ candidates is based on the variables $\de~=~E^{\cms}_{B}-
E^{\cms}_{{\rm beam}}$, the energy difference between the signal $B$ candidate and beam in 
the $\cms$ frame, and $\mbc~=~\sqrt{\left(E^{\cms}_{{\rm beam}}/c^2\right)^2-\left(p^{\cms}
_{B}/c\right)^2}$, the beam-energy constrained mass of the signal $B$ candidate.  The 
candidates satisfying $|\de|<0.3\gev$ and $\mbc>5.2\gevcsq$ are retained for further analysis.

The vertex-constrained kinematic fit is applied to the signal and tagging $B$ candidates and 
to the \dn candidates.  We require $\chi^2/\ndf<500$ for the vertex-constrained fit of the \dn 
meson candidates, where \ndf denotes the number of degrees of freedom.

When $h^0$ is a $\pi^0$ or \etagg candidate, the \bdsth decay has no charged particle originating 
from the primary $B$ decay vertex.  In this case, the $B$ decay vertex is determined by projecting 
the \dn-candidate trajectory onto the beam-interaction profile.  The estimated longitudial resolution 
$\sigma_z$ of a such vertex, obtained from the fit, is required to be less than $0.5$~mm.  This 
requirement is also imposed on the tagging $B$-decay vertices obtained by projecting a single track 
onto the beam interaction profile.

The vertex-constrained kinematic fit for other signal $B$ decay modes requires 
that the $D$ candidate trajectory and the two tracks from the $h^0$ decay originate 
from a common vertex and applies the Gaussian constraints on the position of this 
vertex based on the geometry of the beam interaction profile.  The requirements 
$\sigma_z<0.2$~mm and $\chi^2/\ndf<50$ for the vertex quality are imposed, 
where $\chi^2/\ndf$ is calculated without taking into account the beam interaction 
profile constraint.  These requirements are also imposed on the tagging $B$
decay vertices reconstructed with more than one track.

The vertex position for the tagging $B$ candidate is determined from the kinematic fit 
of well-reconstructed tracks that are not assigned to the signal $B$ candidate decay 
chain~\cite{tagvtx}.

The momentum of the $\pi^0$, \ks, and \etagg candidates, with the invariant mass 
constrained to its nominal value~\cite{pdg}, is used to improve the \de resolution.  
The momenta of the \dn daughters obtained by a mass-constraint fit to the \dn 
candidate are used to calculate the Dalitz variables.  

The continuum background arising from $e^+e^-\to q\bar q$ (where $q=u,~d,~s,~c$) events 
is suppressed with the procedure described in~Refs.~\cite{SFW,KSFW} and 
with the BDT~\cite{bdt1,bdt2} algorithm implemented within the TMVA~\cite{tmva} 
package.  

The $b$ flavor of the tagging $B$ meson is identified from inclusive properties 
of particles that are not associated with the signal $B$ candidate~\cite{TaggingNIM}.  
The tagging information is represented by two parameters: the $b$-flavor charge $q$ 
and the purity $r$.  The parameter $r$ is an event-by-event, MC-determined flavor-tagging 
dilution factor that ranges from $r=0$ for no flavor discrimination to $r=1$ for 
unambiguous flavor assignment.  The data are sorted into seven intervals of $r$.  
For events with $r>0.1$, the wrong tag fractions for six $r$ intervals, $w_l$ ($l=1,2,\dots,6$), 
and their differences between \bn and \bnbar decays, $\Delta w_l$, are determined from 
semileptonic and hadronic $b\to c$ decays~\cite{Tagging}.  If $r\leq0.1$, the wrong tag 
fraction is set to $0.5$ and the tagging information is not used.  The total effective 
tagging efficiency, $\varepsilon_{\rm eff} = \sum(f_l\times(1-2w_l)^2)$, is $0.3$, where 
$f_l$ is the fraction of events in the category $l$.  The parameter $\qb = q_B\left(1-2w\right)
/(1-q_{B}\Delta w)$ is used instead of the parameter $q_B$, defined in Eq.~(\ref{eq:master-formula}), 
to account for the wrong tag.

The signal yields of \bdsth modes are obtained from an extended unbinned maximum 
likelihood fit of the \dembc two dimensional distribution in the region 
$\de\,\in\,(-0.15\gev,~0.30\gev)\cap\mbc\,\in\,(5.20\gevcsq,~5.29\gevcsq)$.  
The signal yield of \bptodpi events is obtained from an extended unbinned maximum 
likelihood fit of the \de distribution in the region $(-0.10\gev,~0.15\gev)$ 
for $\mbc~\in~(5.272\gevcsq,~5.287\gevcsq)$.

The sideband region is defined as the union of two rectangular regions in the 
\dembc plane: $\mbc~\in~(5.23\gevcsq,~5.26\gevcsq)\ \cap\ \de\ \in\ (-0.15\gev,
~0.30\gev)$ and $\mbc~\in~(5.26\gevcsq,~5.29\gevcsq)\ \cap\ \de\ \in\ (0.12\gev,~0.30\gev)$.

The selection criteria and the analysis procedure are tested using the Monte Carlo (MC) 
simulation and fixed before performing the fit of the \cpconj violation parameters.  
The MC events are generated with \verb@EvtGen@~\cite{evtgen}.  Final-state radiation 
from charged particles is simulated during the event generation using \verb@PHOTOS@~\cite{photos}.  
The generated events are processed through the detailed detector simulation based on 
\verb@GEANT3@~\cite{geant}.

\section{\boldmath \bptodpi sample}\label{sec:bdpip}
The \bptodpi control sample is experimentally clean and has kinematic properties 
and detection efficiency similar to the \bdsth decay. We use this process to 
select a sample of $D$ mesons in the flavor eigenstate and to measure the 
parameters \ki defined in~Eq.~(\ref{eq:k}).

\subsection{Signal yield}
Three components are included in the fit of the \de distribution: 
signal, \bptodk background and combinatorial background.

\begin{figure}[htb]
\includegraphics[width=0.5\textwidth]{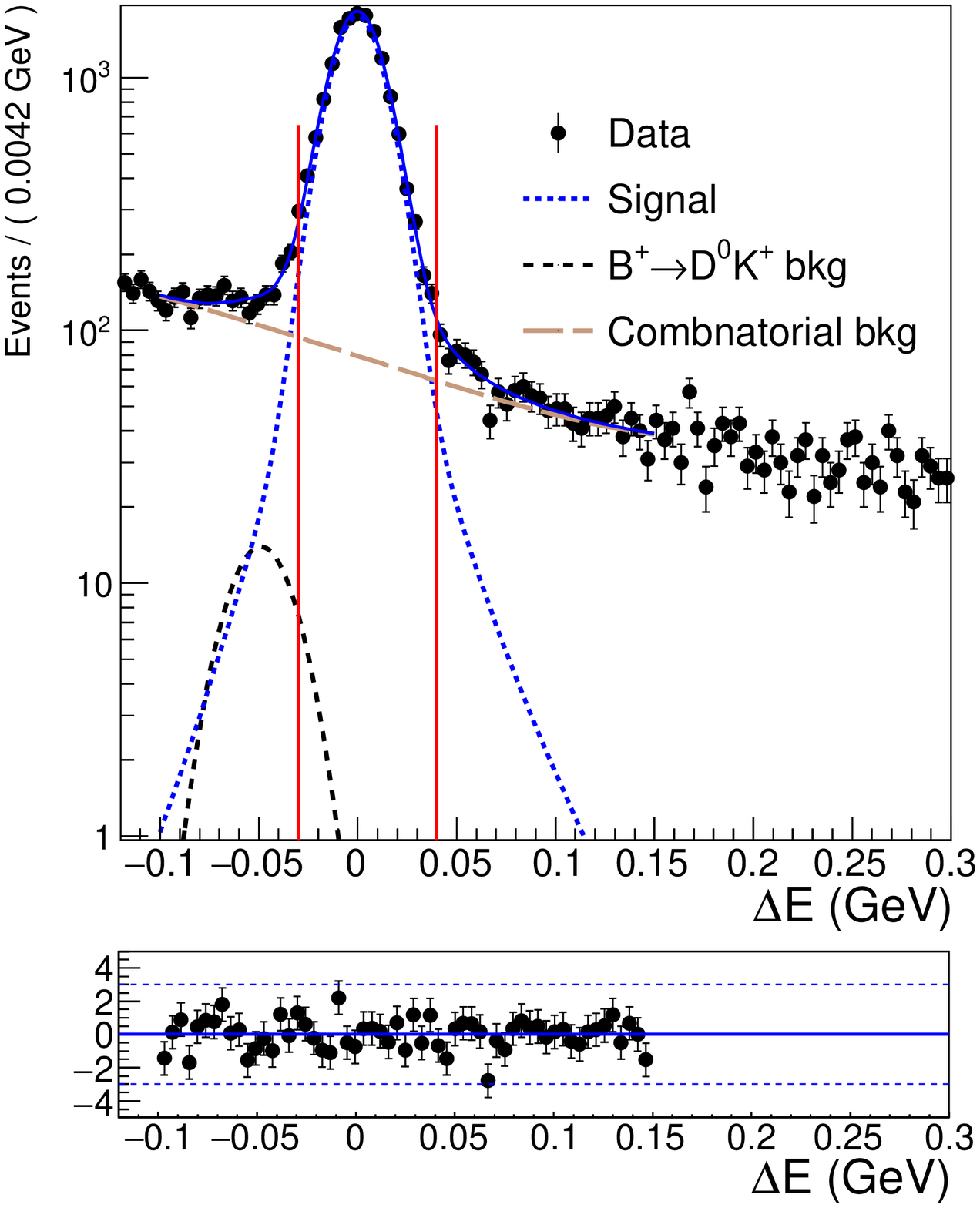}
\caption{\de distribution for \bptodpi candidates.  Black circles with error bars show 
data, the solid blue line is the complete fit function, the dashed blue line is the signal 
component, the dashed black line is the background from \bptodk decays, the dashed 
brown line is the combinatorial background.  Vertical red lines show the signal area.  
Histogram with the pulls of the data with respect to the fit curve is shown at the bottom 
(with horizontal blue dashed lines at pull values of $\pm3$).}
\label{fig:de-bptodpi}
\end{figure}

The signal distribution is parameterized by the sum of a Gaussian and two 
Crystal Ball functions~\cite{CB} with a common peak position.  The mean and the 
Gaussian width are free fit parameters while the other parameters are 
fixed to the values obtained from simulation.  Background from 
the \bptodk decays is parameterized by a Gaussian function with all 
parameters fixed from simulation.  Combinatorial background is parameterized 
by a second-order Chebyshev polynomial.  The parameters of the 
combinatorial background shape are obtained from the fit.  The \de 
distribution for \bptodpi candidates and the results of the fit are 
shown in Fig.~\ref{fig:de-bptodpi}. Yields of the signal and background 
components are listed in Table~\ref{tab:bp2d0pi_de_fit}.

\begin{figure*}[htb]
\subfloat[]{\label{fig:bp_dp_sig}\includegraphics[width=0.45\textwidth]{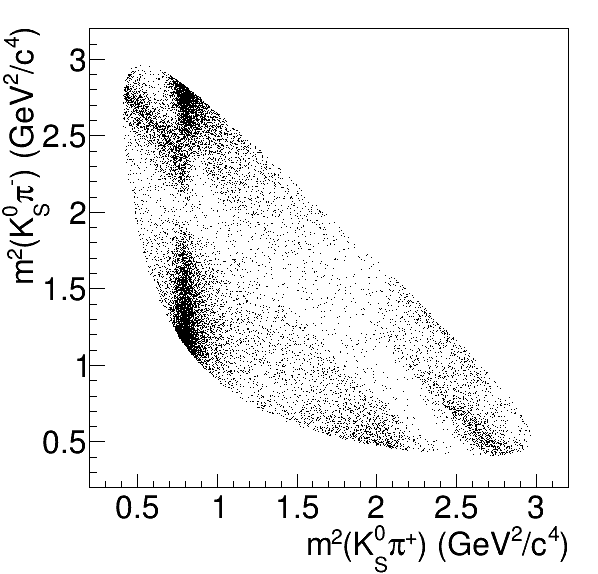}}\hfill
\subfloat[]{\label{fig:bp_dp_bkg}\includegraphics[width=0.45\textwidth]{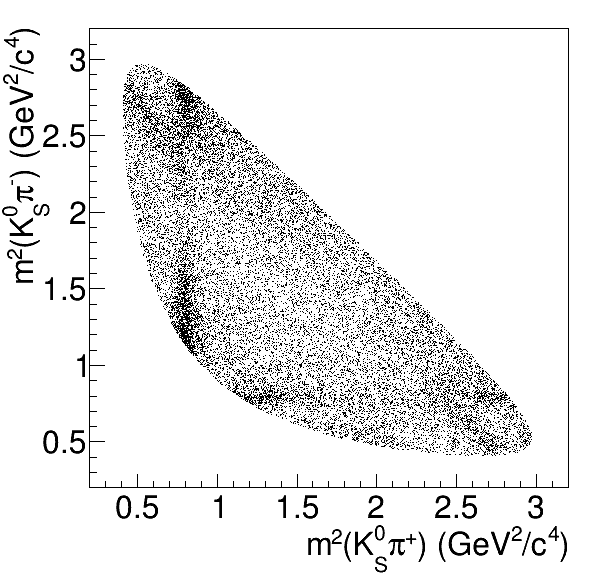}}
\caption{Dalitz plot distributions for \dkpp candidates with $D$ from 
\bptodpi decay in the signal (a) and sideband~(b) areas.}
\label{fig:bp-dalitz}
\end{figure*}

\begin{table}[htb]
\caption{Fit results of the \de distribution for \bptodpi candidates.  
The numbers of events and the  fraction of signal events are shown for the 
signal \de region.}
\label{tab:bp2d0pi_de_fit}
\begin{tabular}
 {@{\hspace{0.5cm}}l@{\hspace{0.5cm}}  @{\hspace{0.5cm}}c@{\hspace{0.5cm}}}
\hline \hline
 {Parameter}               & {Value} \\ \hline
  Signal yield             & $(1.375 \pm 0.014)\times 10^{4}$ \\
  \bptodk yield            & $18.7   \pm 9.8$                 \\
  Combinatorial bkg. yield & $1295   \pm  79$                 \\
  Signal fraction ($\%$)   & $91.3   \pm 0.9$                 \\
  \hline \hline
\end{tabular}
\end{table}

The parameters \ki are measured using the events in the \de interval 
between $-30$ and $40\mev$.  This interval is optimized to suppress 
the background from \bptodk events without significant signal-efficiency loss.

\subsection{\boldmath Measurement of parameters \ki}\label{sec:K_measurement}
The charged pion from the \bptodpi decay tags the flavor of the $D$ meson.  
Therefore, the fraction of the signal events corresponding to the 
$i$-th Dalitz plot bin equals~\ki.

The Dalitz distribution for \dkpp in the signal \de range, where 
the $D$ meson is produced in \bptodpi decays, is shown in~Fig.~\ref{fig:bp_dp_sig}.  
The fraction of signal, $f\subsig=(91.3\pm 0.9)\%$, is obtained from a fit 
of the \de distribution.  The Dalitz plot for events from the \de-\mbc sideband 
is shown in Fig.~\ref{fig:bp_dp_bkg}.  The binned background distribution 
is obtained from this data.

\begin{table}[htb]
 \caption{The values of the parameters \ki measured with the \bptodpi data sample.  
 The values are not corrected for the detection efficiency.}
 \label{tab:Kmeasured}
 \begin{tabular}
  { @{\hspace{0.7cm}}c@{\hspace{0.7cm}} @{\hspace{0.7cm}}r@{\hspace{0.7cm}} @{\hspace{0.7cm}}r@{\hspace{0.7cm}}}
  \hline\hline
  Bin ($i$) & \multicolumn{1}{l}{\ki (\%)} & \multicolumn{1}{l}{\kmi (\%)}\\ \hline
  $1$ & $17.42\pm0.32$ & $7.81\pm0.25$ \\
  $2$ & $ 7.51\pm0.22$ & $1.29\pm0.10$ \\
  $3$ & $10.24\pm0.26$ & $2.58\pm0.14$ \\
  $4$ & $ 2.85\pm0.14$ & $1.16\pm0.10$ \\
  $5$ & $ 9.45\pm0.25$ & $4.25\pm0.17$ \\
  $6$ & $ 7.31\pm0.22$ & $1.73\pm0.11$ \\
  $7$ & $10.48\pm0.26$ & $1.18\pm0.10$ \\
  $8$ & $12.46\pm0.28$ & $2.38\pm0.14$ \\ \hline
  \hline
 \end{tabular}
\end{table}

The values of the parameters \ki are listed in Table~\ref{tab:Kmeasured}.  
The uncertainties shown include the statistical uncertainty of the signal 
sample and the uncertainty due to background evaluation, added in quadrature.  
The systematic uncertainties associated with the background Dalitz plot 
distribution are neglected because the background fraction is very small.

\section{\boldmath \bdsth sample}
\subsection{Background components}
Three background components are considered for the \bdsth candidates:
\begin{itemize}
 \item combinatorial background from non-resonant light quark production (continuum background);
 \item combinatorial background from \bbbar events; and
 \item background from partially reconstructed $B$ decays.
\end{itemize}
Background from partially reconstructed decays is dominated by $B\to\dnbar\rho$ and 
$B\to\dbar{}^{*}\pi^0$ for the \bdpi mode and by $B\to\dbar{}^*\rho$ for the \btodstpi mode.  
These processes, reconstructed with one missing pion, lead to a concentration below 
$-0.1\gev$ in the \de distribution.  The background in all other channels is dominated by the 
combinatorial contribution with featureless \de distribution.

The background contribution from charmless \bn decays is suppressed by requiring the presence 
of a \dn candidate and thus is found to be negligible in this measurement.

\subsection{Signal yield}
A two-dimensional unbinned maximum likelihood fit of the \dembc distribution is 
performed for each signal mode.  The probability density function (PDF) contains 
four components, corresponding to the signal and three backgrounds introduced above.

The signal \de distributions are parameterized by the sum of a Gaussian and two Crystal 
Ball functions with a common peak position.  The signal \mbc distributions are parameterized 
by a function introduced in Ref.~\cite{NskFcn} and referred to as the Novosibirsk distribution.  
The peak position in the \dembc plane is obtained  from the fit while the other parameters are 
fixed at the values obtained from simulation.

The \de distributions for events from continuum background are parameterized by a second-order 
Chebyshev polynomial.  The \de distributions of the combinatorial background from the \bbbar 
events are parameterized by an exponential function.  The \mbc distributions of the combinatorial 
backgrounds are parameterized by an ARGUS function~\cite{argus}.  The parameters of the \de PDF are 
obtained from the fit while those of the \mbc PDF are fixed at the values obtained from simulation.

\begin{figure*}[htb]
\subfloat[]{\label{fig:de_pi0}   \includegraphics[width=0.49\textwidth]{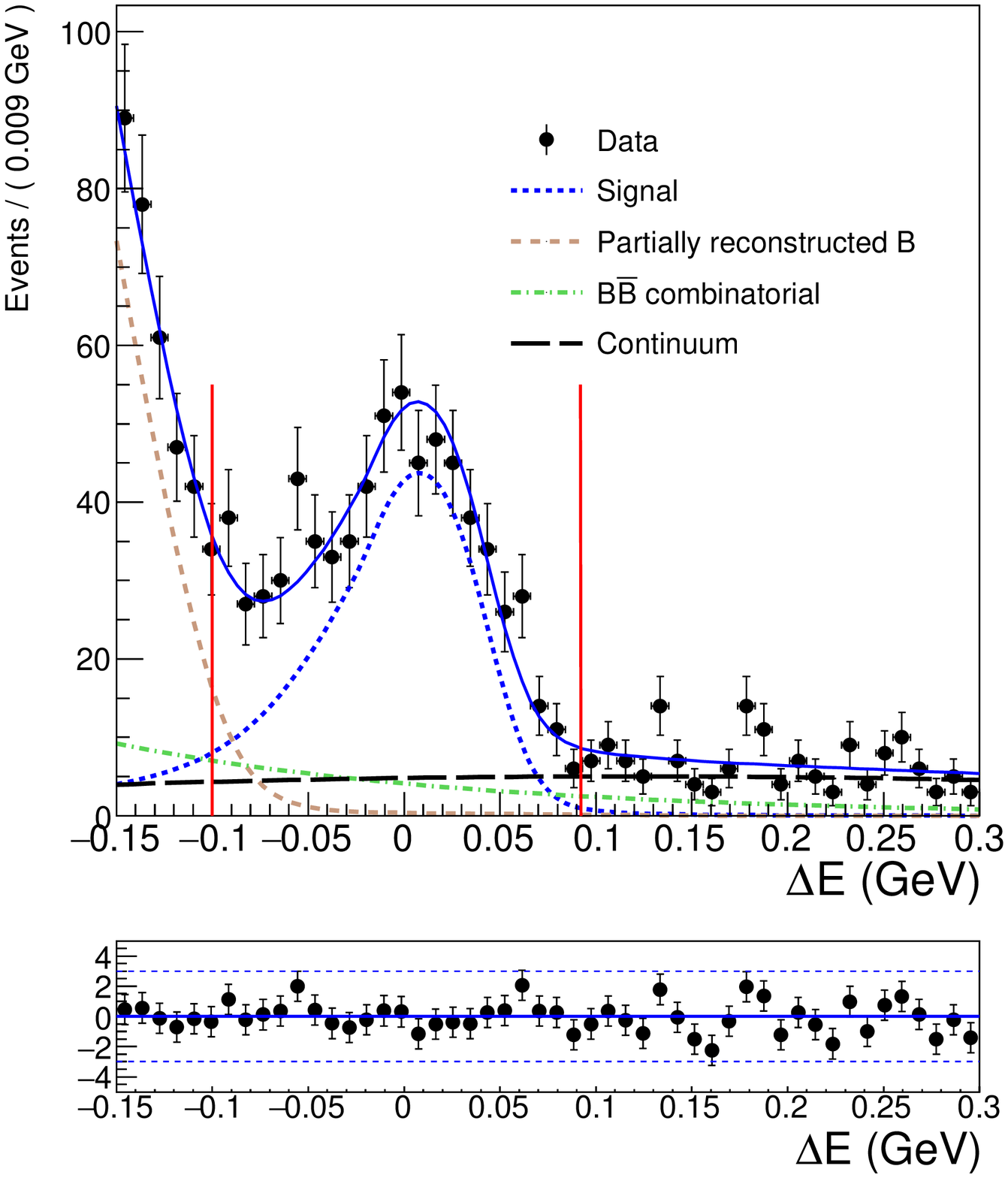}}\hfill
\subfloat[]{\label{fig:mbc_pi0}  \includegraphics[width=0.49\textwidth]{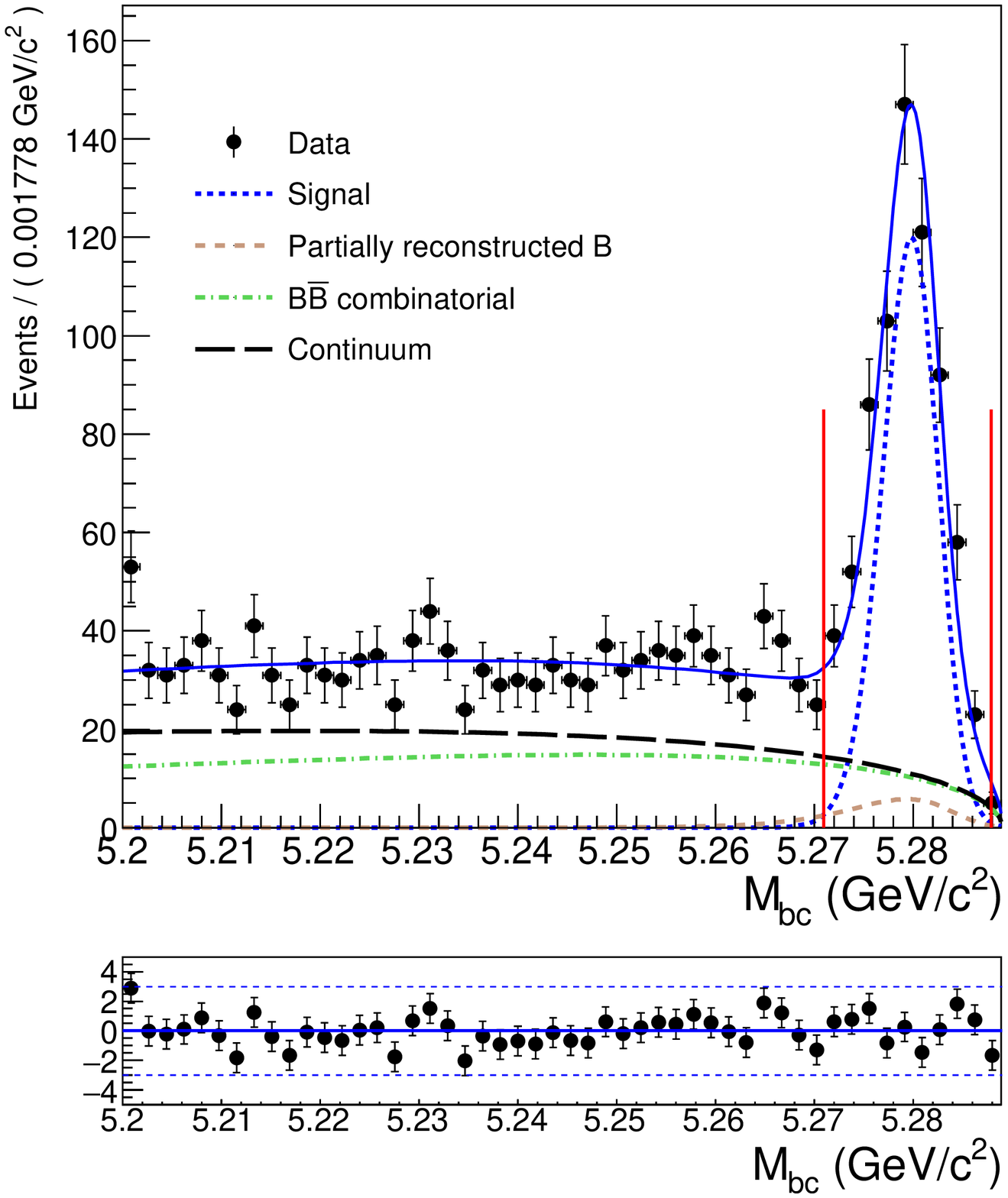}}\\
\subfloat[]{\label{fig:de_omega} \includegraphics[width=0.49\textwidth]{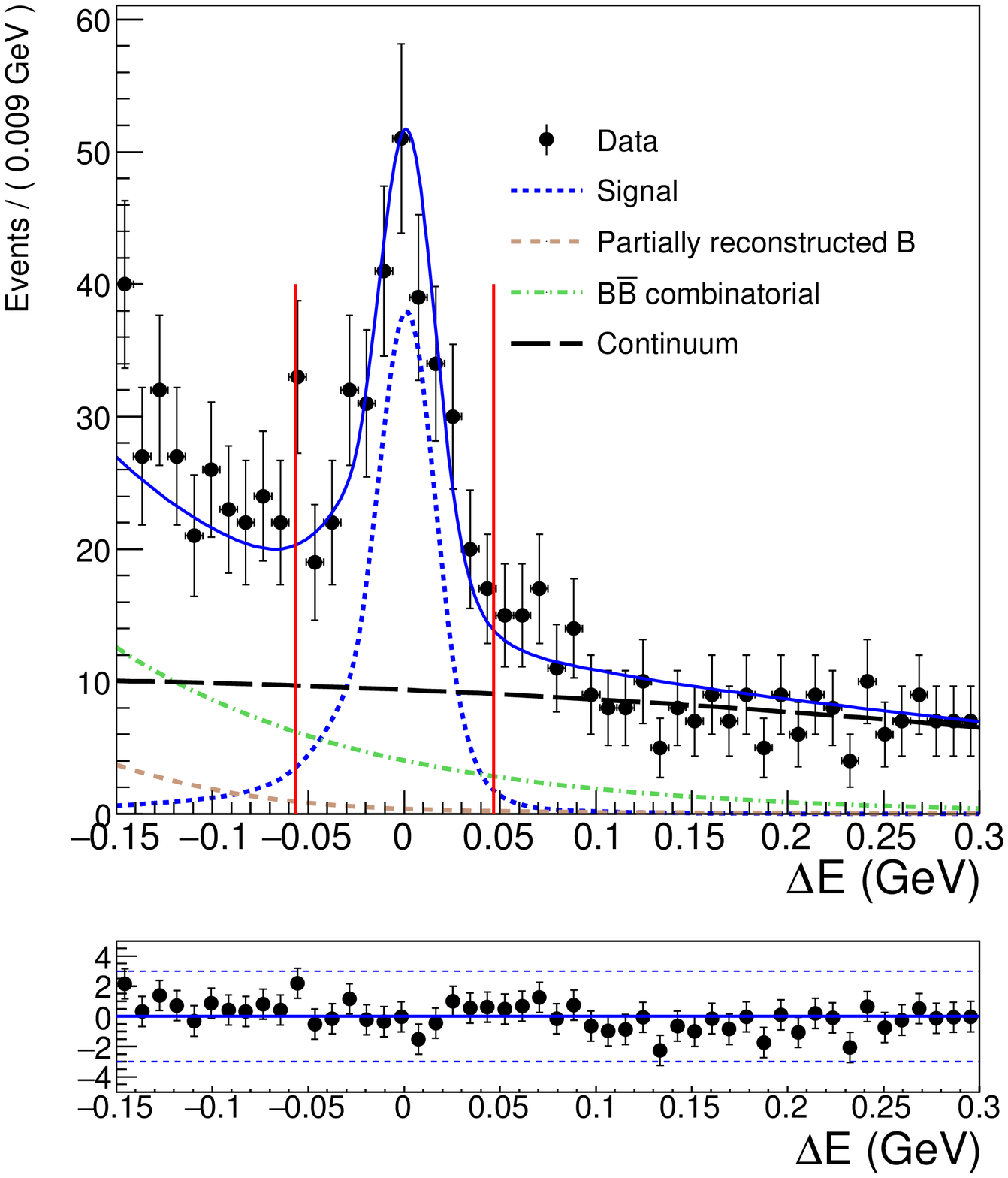}}\hfill
\subfloat[]{\label{fig:mbc_omega}\includegraphics[width=0.49\textwidth]{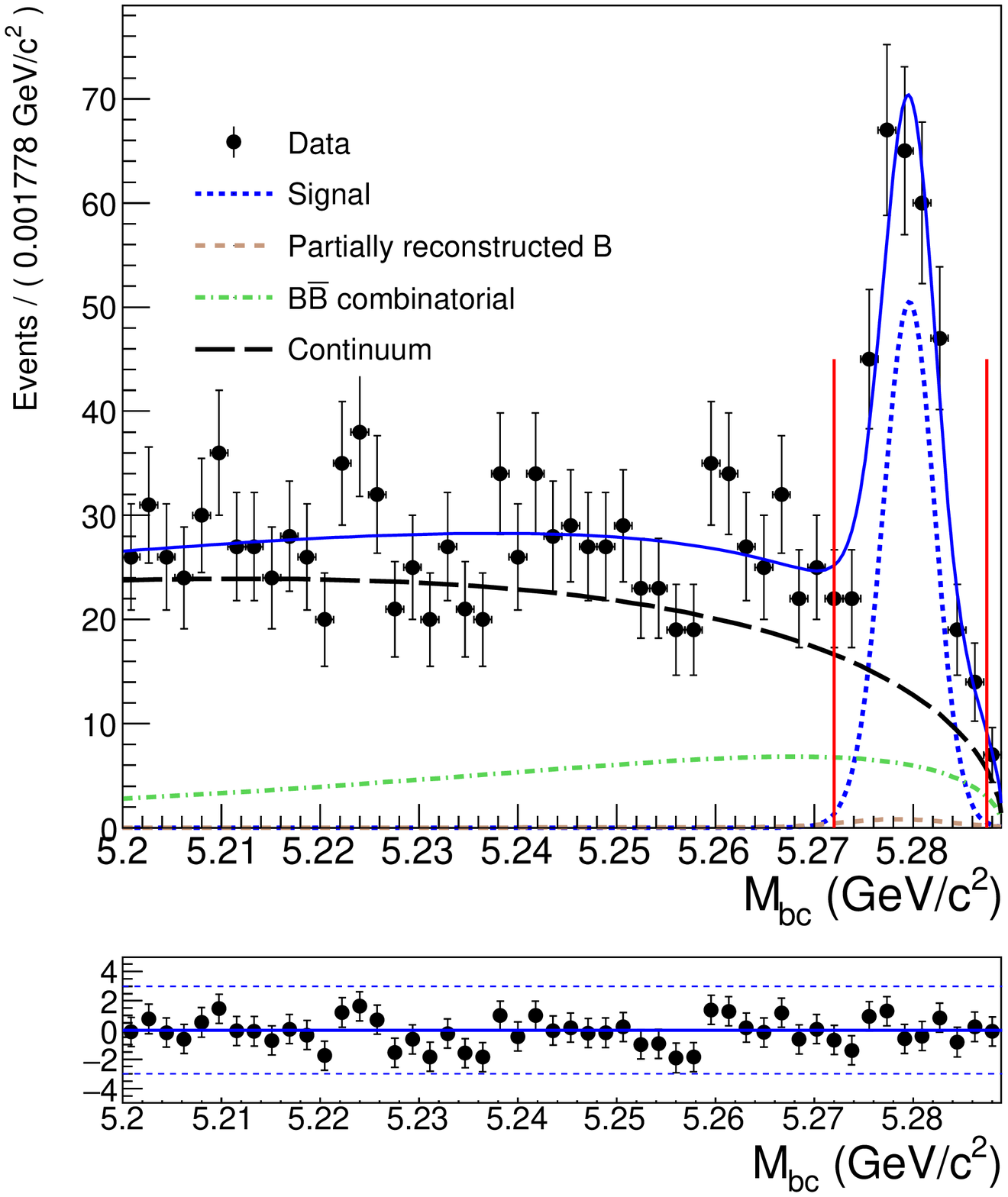}}
\caption{\de fit projections for the signal \mbc regions (a,\,c) and \mbc fit 
projections for the signal \de regions (b,\,d) for the \bdpi (a,\,b) and \bdomega (c,\,d)
candidates.  Black circles with errors show data, continuous blue lines show projections of 
complete fit functions, dashed blue lines show signal components, dashed black lines show 
continuum background components, dashed brown lines show background from partially reconstructed 
$B$ decays and dot-dashed lines show combinatorial background from \bbbar events.  Histograms with 
the pulls of the data with respect to the fit curves are shown at the bottom of each plot (with 
horizontal blue dashed lines at pull values of~$\pm3$).}
\label{fig:de-mbc}
\end{figure*}

\begin{figure*}[htb]
\subfloat[]{\label{fig:de_etagg}  \includegraphics[width=0.193\textwidth]{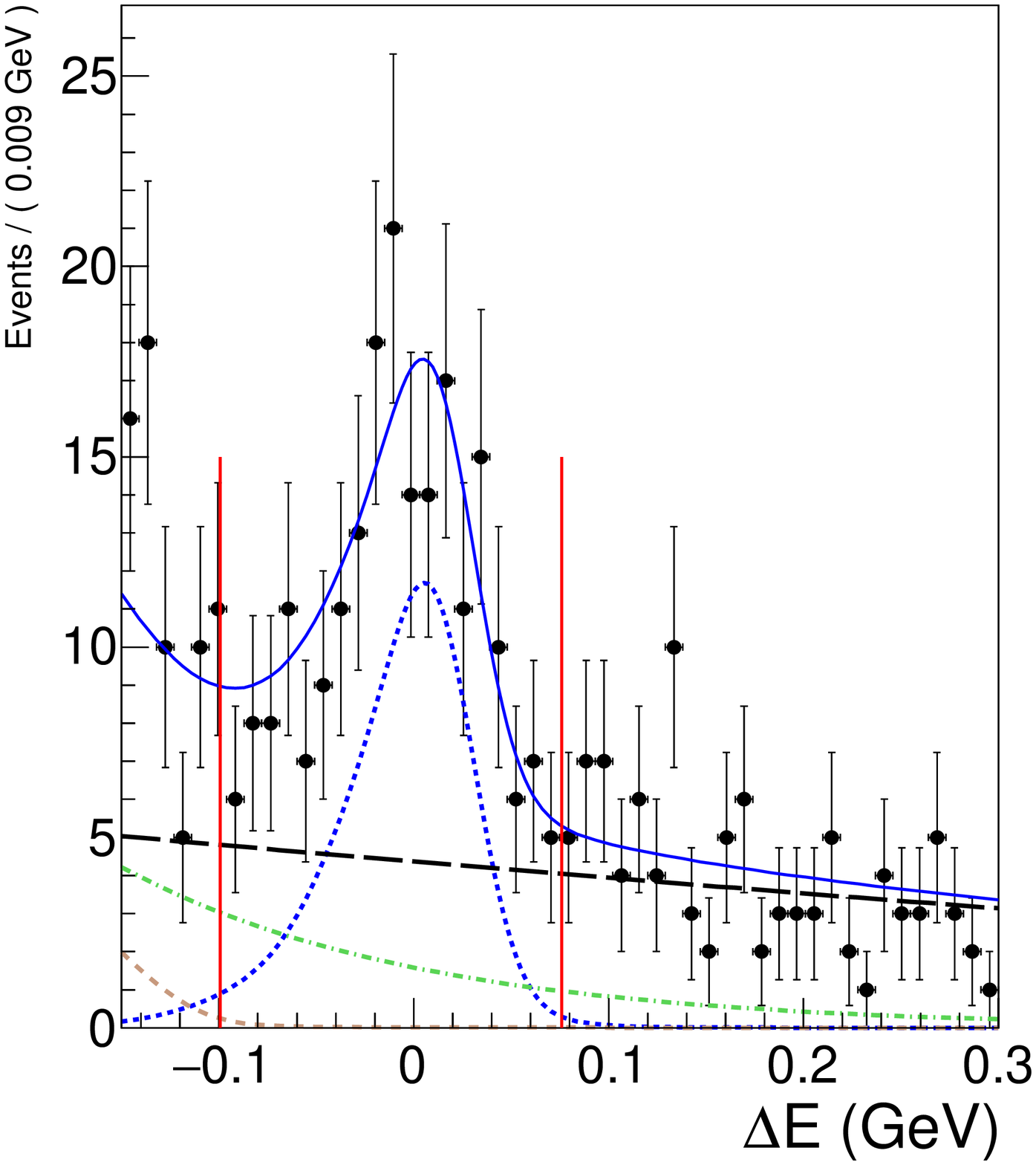}}\hfill
\subfloat[]{\label{fig:de_etapp}  \includegraphics[width=0.193\textwidth]{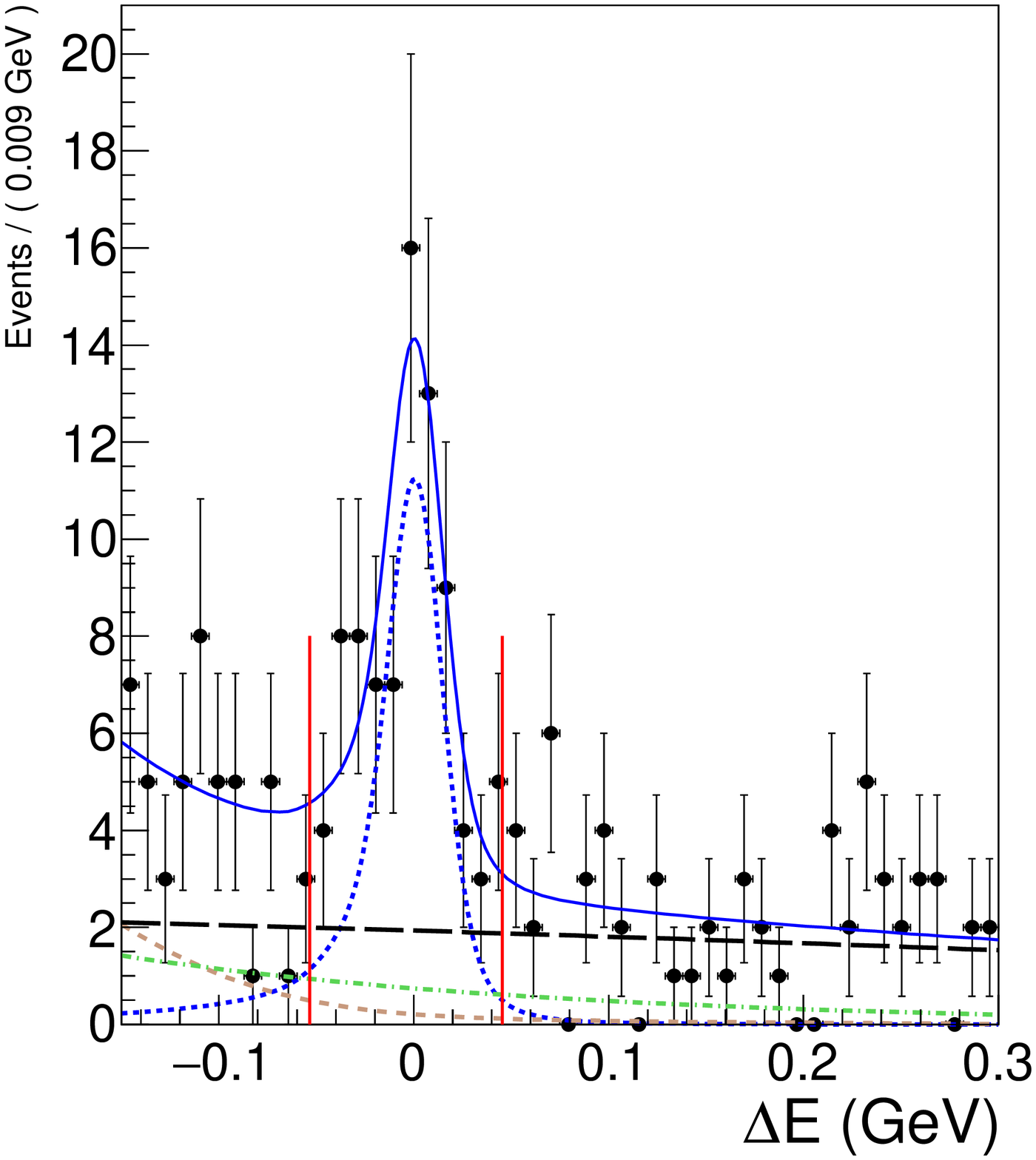}}\hfill
\subfloat[]{\label{fig:de_etap}   \includegraphics[width=0.193\textwidth]{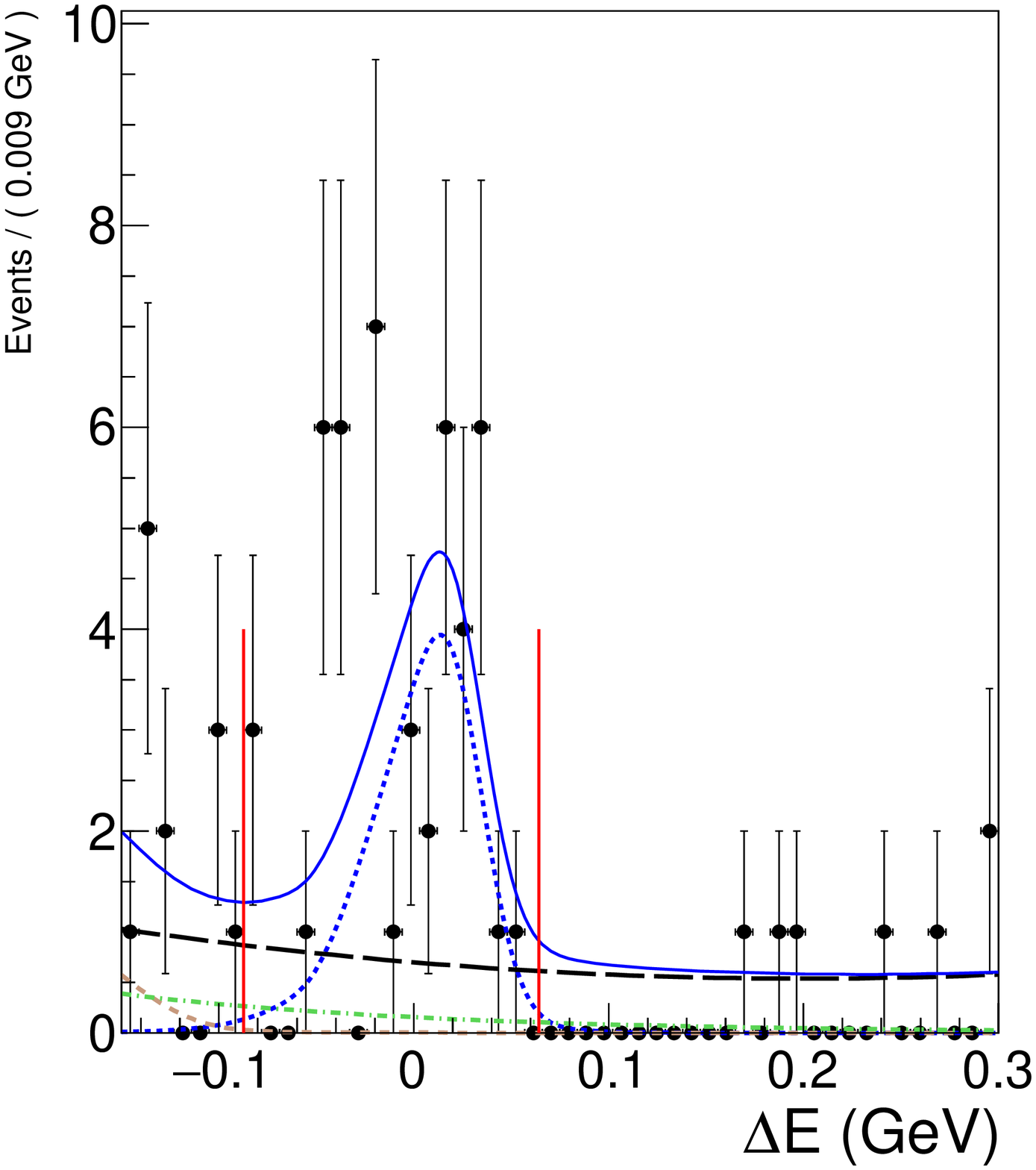}}\hfill
\subfloat[]{\label{fig:de_dstpi}  \includegraphics[width=0.193\textwidth]{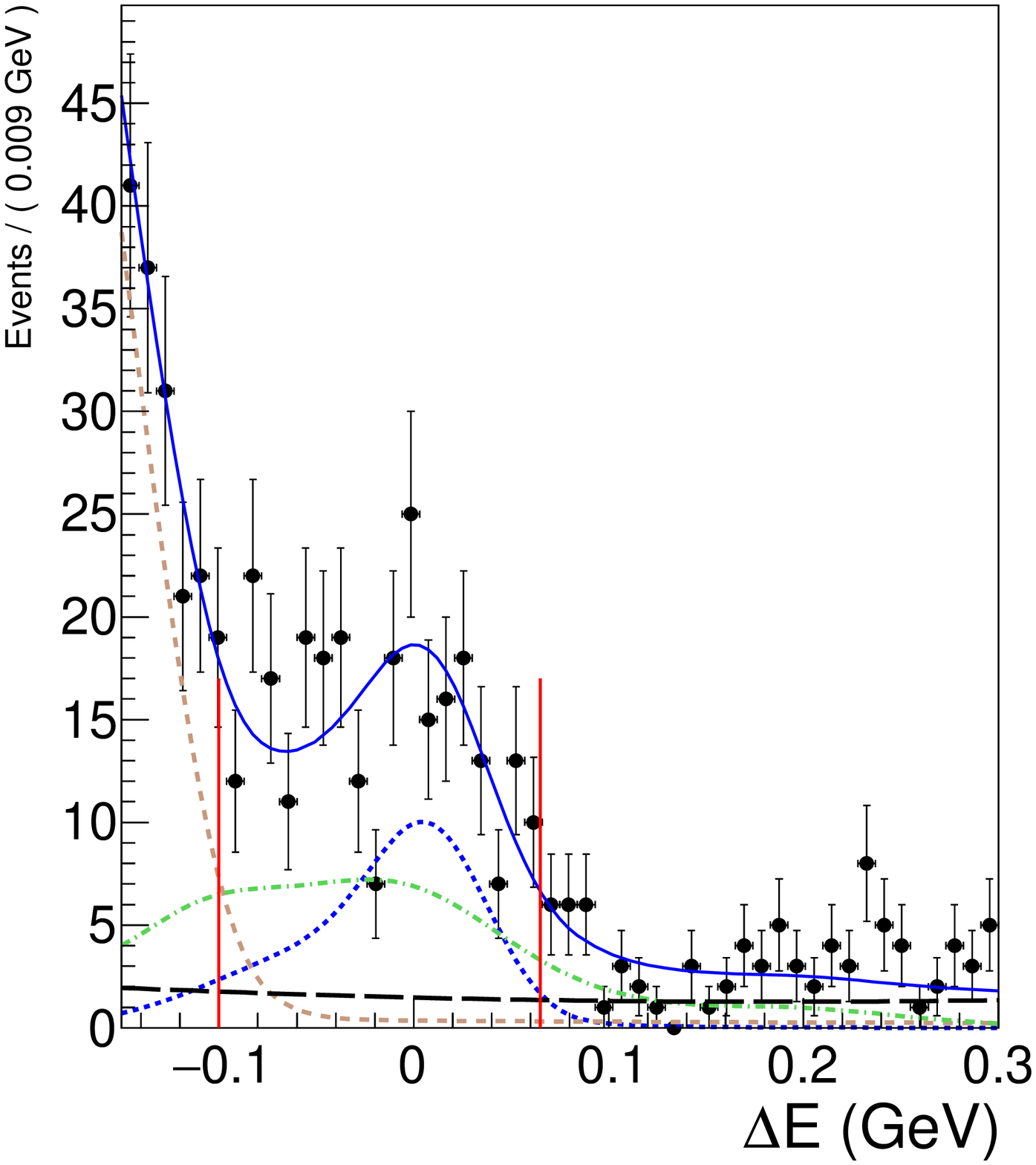}}\hfill
\subfloat[]{\label{fig:de_dsteta} \includegraphics[width=0.193\textwidth]{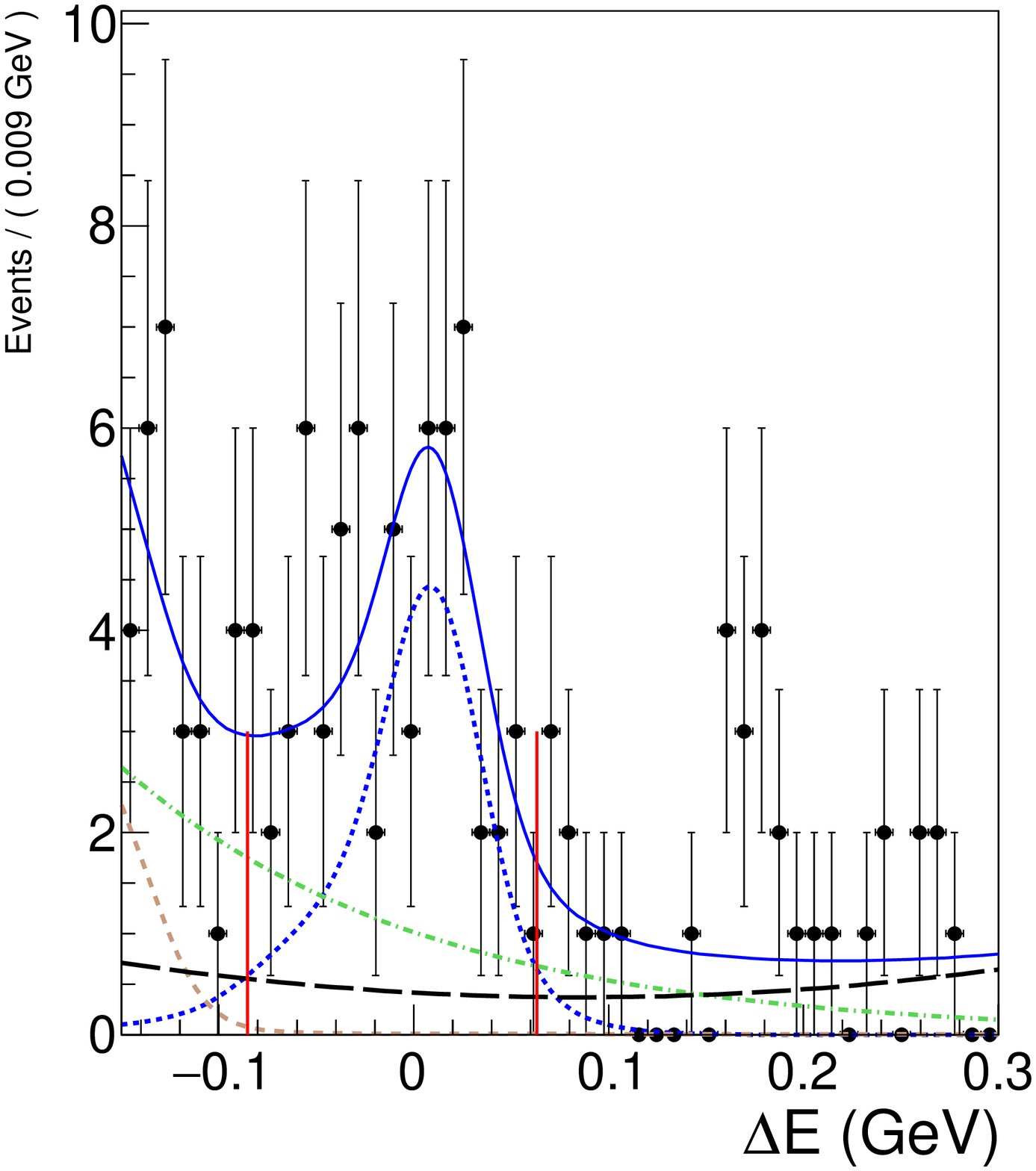}}\\
\subfloat[]{\label{fig:mbc_etagg} \includegraphics[width=0.193\textwidth]{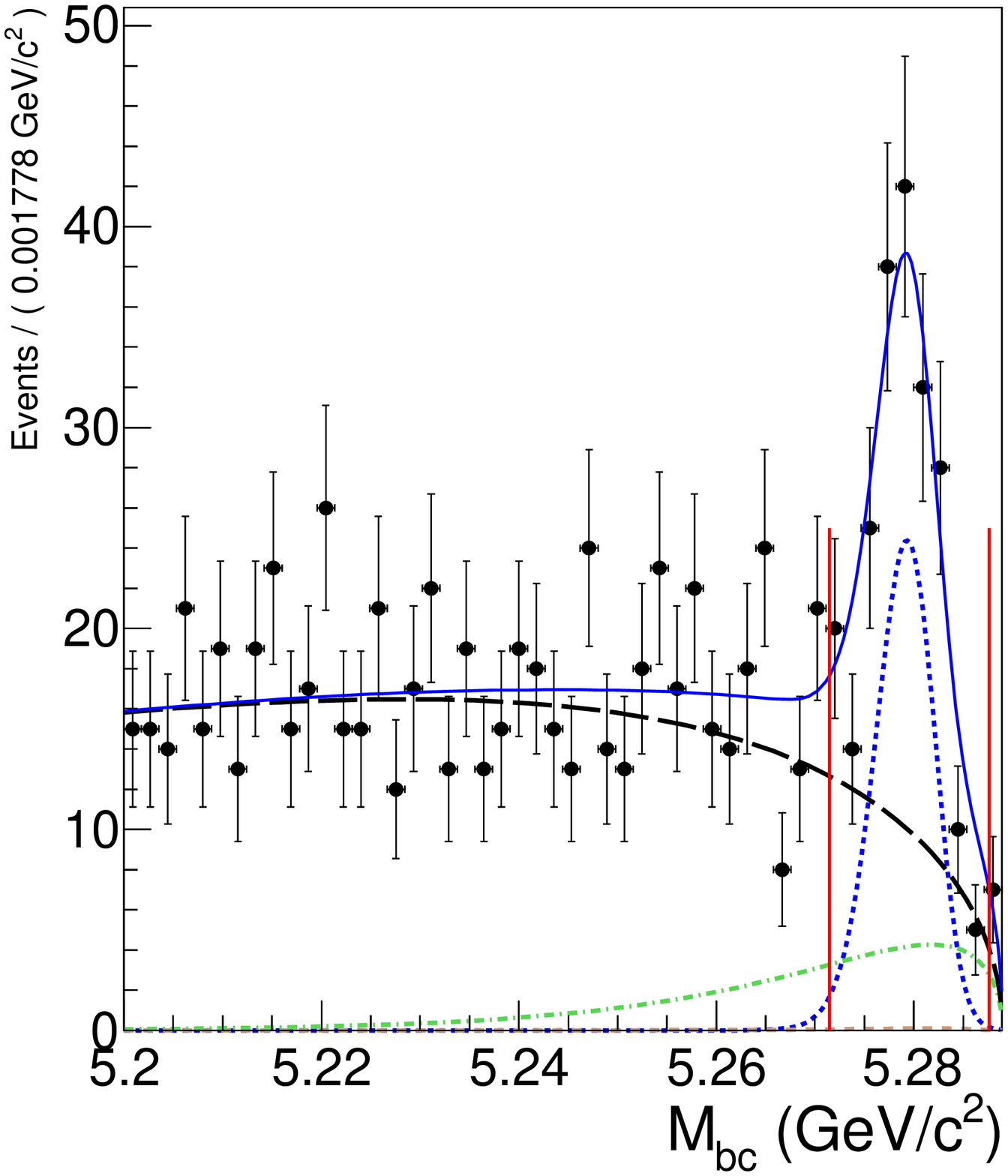}}\hfill
\subfloat[]{\label{fig:mbc_etapp} \includegraphics[width=0.193\textwidth]{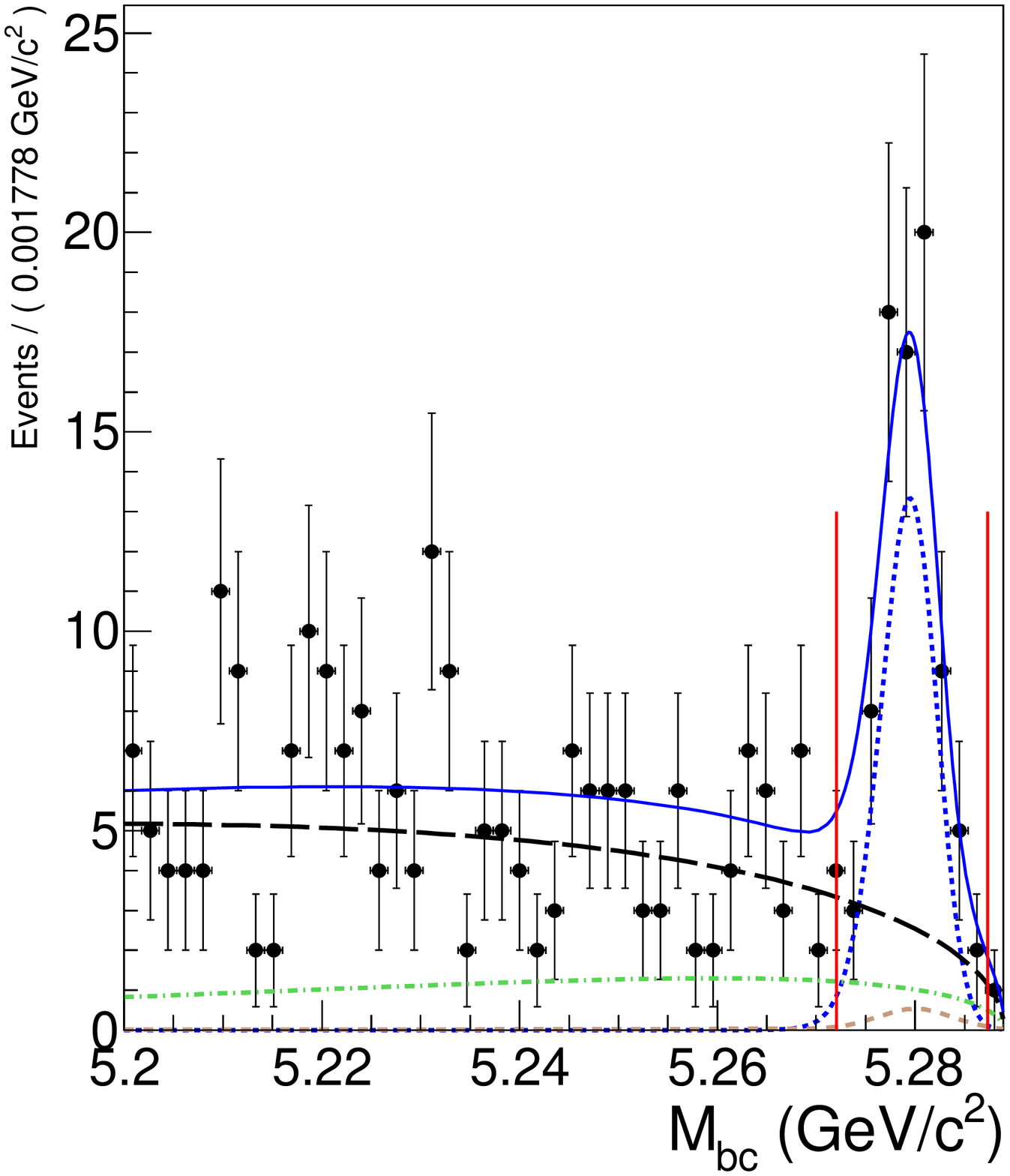}}\hfill
\subfloat[]{\label{fig:mbc_etap}  \includegraphics[width=0.193\textwidth]{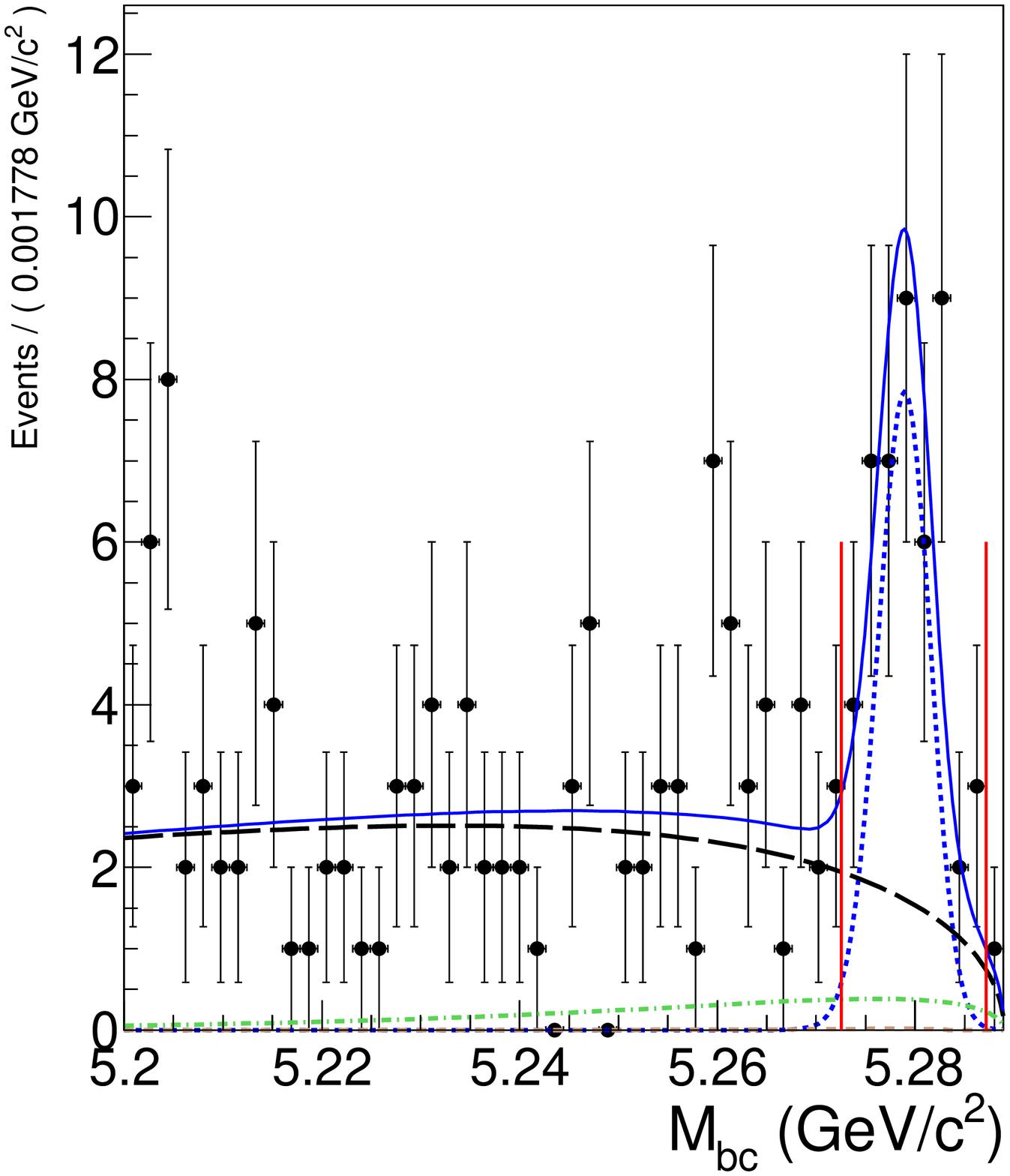}}\hfill
\subfloat[]{\label{fig:mbc_dstpi} \includegraphics[width=0.193\textwidth]{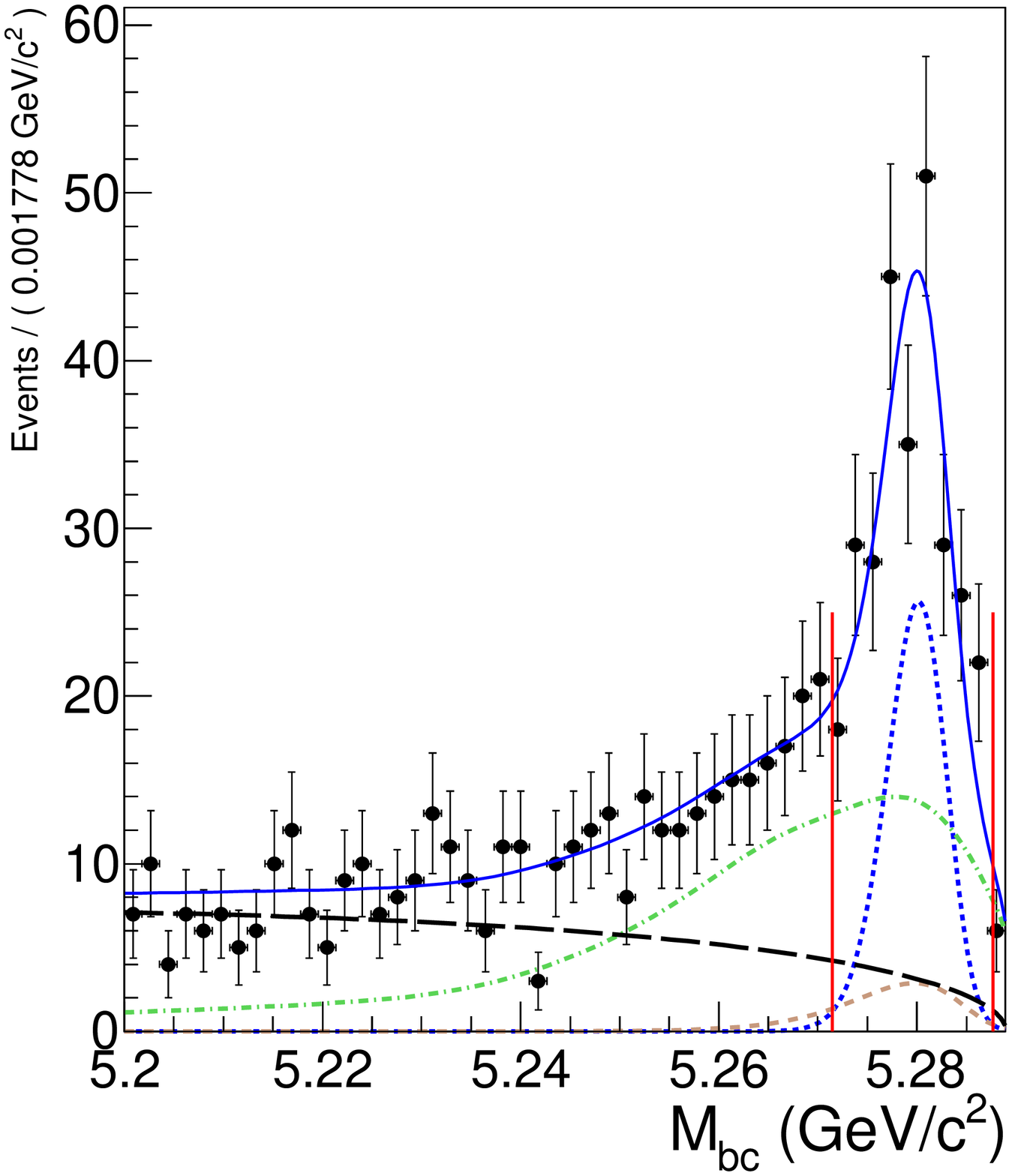}}\hfill
\subfloat[]{\label{fig:mbc_dsteta}\includegraphics[width=0.193\textwidth]{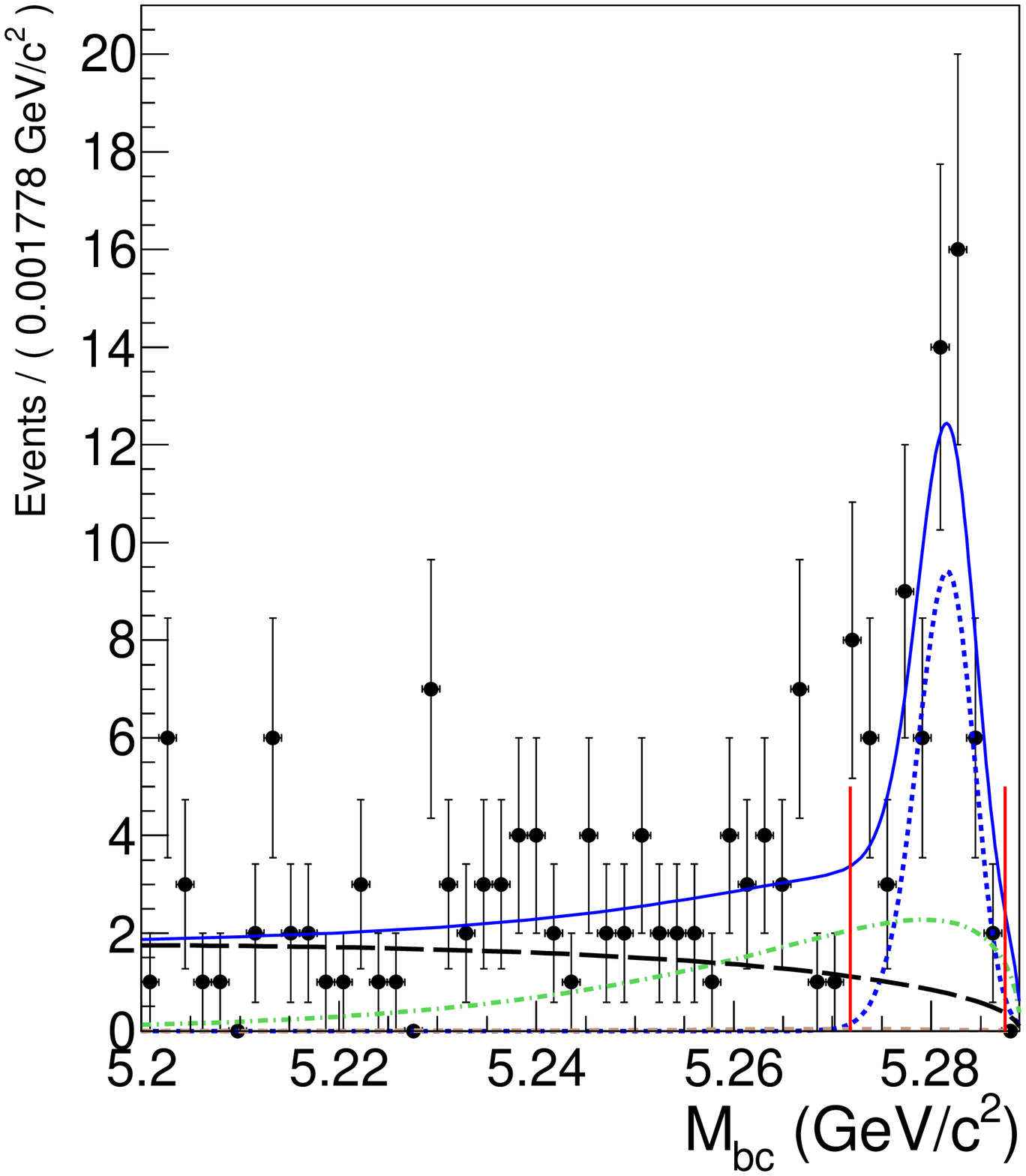}}\\
\caption{\de fit projections for the signal \mbc regions (a\,--\,e) and \mbc fit projections for the 
signal \de regions (f\,--\,j) for the \bdeta, \etagg (a,\,f), \bdeta, \etappp (b,\,g), \bdetap (c,\,h), 
\btodstpi (d,\,i), and \btodsteta (e,\,j) candidates.  Black circles with errors show data, continuous 
blue lines show projections of complete fit functions, dashed blue lines show signal components, dashed 
black lines show continuum background components, dashed brown lines show background from partially 
reconstructed $B$ decays and dot-dashed lines show combinatorial background from \bbbar events.}
\label{fig:de-mbc_other}
\end{figure*}

The \de distributions of the background from partially reconstructed $B$ decays are 
parameterized by the following function:
\begin{equation}\label{eq:prt_de_shape}
\begin{split}
 p_{\rm pr}&(\de) \propto 1+\zeta_{\rm l}\left(\de-\de_0\right)\\
 &+s\ln\left(1+b \exp\left[{\frac{\left(\zeta_{\rm r}-\zeta_{\rm l}\right)\left(\de-\de_0\right)}{s}}\right]\right).
\end{split}
\end{equation}
This function describes two asymptotically straight lines smoothly merged near the point given 
by the $\de_0$ parameter whose slopes are given by~$\zeta_{\{{\rm r,l}\}}$.  The parameter $s$ 
determines the curvature at the junction.  If the $B$ candidate decay chain contains a $\pi^0$ 
or $\eta$ reconstructed in the \gaga final state, the \mbc distribution of the background from 
partially reconstructed $B$ decays is parameterized by the Novosibirsk function; otherwise, it 
is parameterized by the sum of ARGUS and Gaussian functions.  All parameters are fixed at the 
values obtained from simulation except for the values of the $\de_0$ parameter for the \bdpi 
and \btodstpi modes that are obtained from the fit.  

Several correlations between the \de and \mbc distributions are taken into 
account.  A left-side tail of the signal \de distribution is due to $\pi^0$ 
or $\eta$ candidate where only one photon was identified correctly. This 
partially wrong combination leads to correlated shift both in \de and \mbc.  
A similar correlation appears in the distributions of the background from 
partially reconstructed $B$ decays.  The width of the signal \de distribution 
for the $B$ candidates with the $\eta$ or $\omega$ reconstructed in the \ppp 
final state is determined by the charged final state particles momentum 
resolution if both final state photons are correctly assigned.  For such 
candidates, the \de and \mbc distributions are correlated.  That correlation 
is accommodated by introducing a \de dependence of the signal \mbc PDF parameters.  
This parameterization is equivalent to a two-dimensional Gaussian function.  
No significant correlation is found for the combinatorial background.  The 
values of parameters required to employ the correlations are obtained from 
simulation.

The fit projections for the \bdpi and \bdomega modes are shown in~Fig.~\ref{fig:de-mbc}.  
The fit projections for the other signal modes are shown in~Fig.~\ref{fig:de-mbc_other}.  
The fractions of background from partially reconstructed $B$ decays are small for all 
modes except \bdpi and \btodstpi (compare the \de distributions below~$-0.1\gev$ for 
\bdpi and \bdomega in Fig.~\ref{fig:de-mbc}, for example) and cannot be determined from 
the fit.  These fractions are fixed relative to the fractions of combinatorial background 
from \bbbar events using the values obtained from MC simulation.

\begin{table}[htb]
\caption{ Results of the \dembc fit for \bdsth data. The numbers of events 
$N\subsig$ and the fractions $f\subsig$ of signal events obtained from 
the fit for the signal \dembc regions are shown.}
\label{tab:data_de_mbc}
\begin{tabular}
  { @{\hspace{0.5cm}}l@{\hspace{0.5cm}} @{\hspace{0.5cm}}c@{\hspace{0.5cm}} @{\hspace{0.5cm}}c@{\hspace{0.5cm}} }
 \hline\hline
 {\bf Mode}  & $N\subsig$     & $f\subsig$ (\%) \\
 \hline
 \bdpi       & $464 \pm26$   & $72.1\pm4.1$ \\
 \bdetagg    & $ 99 \pm14$   & $50.5\pm7.0$ \\
 \bdetappp   & $51.3\pm8.8$  & $66  \pm11$  \\
 \bdomega    & $182 \pm18$   & $58.4\pm5.7$ \\
 \bdetap     & $28.2\pm6.4$  & $70  \pm16$  \\
 \btodstpi   & $103 \pm17$   & $44.1\pm7.4$ \\
 \btodsteta  & $36.1\pm7.6$  & $64  \pm13$  \\
 Total       & $962\pm41$    & $61  \pm2.6$ \\
 \hline\hline
 \end{tabular}
\end{table}

\begin{figure*}[htb]
\subfloat[]{\label{fig:dp_bdh_b0} \includegraphics[width=0.45\textwidth]{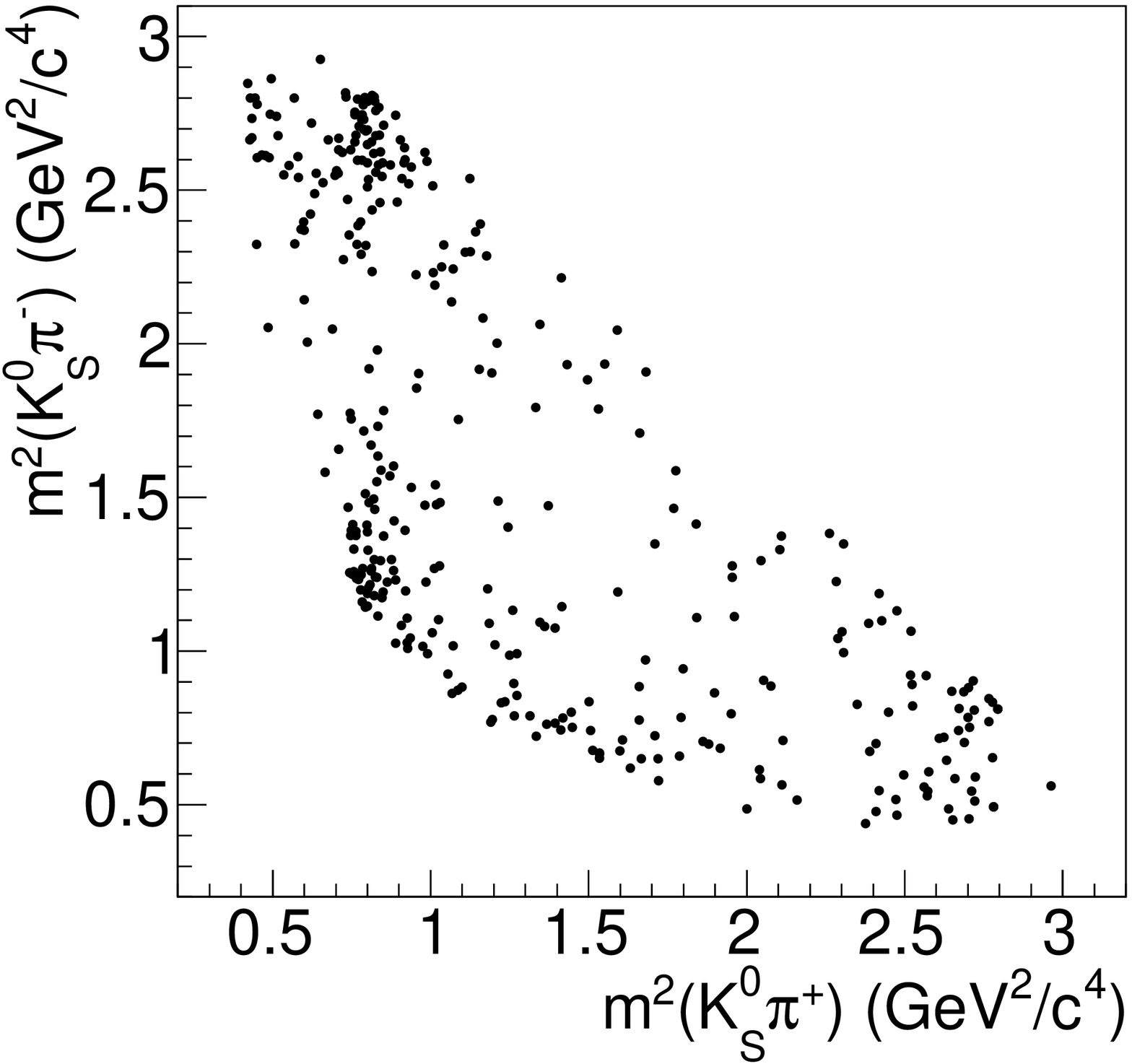}}\hfill
\subfloat[]{\label{fig:dp_bdh_b0b}\includegraphics[width=0.45\textwidth]{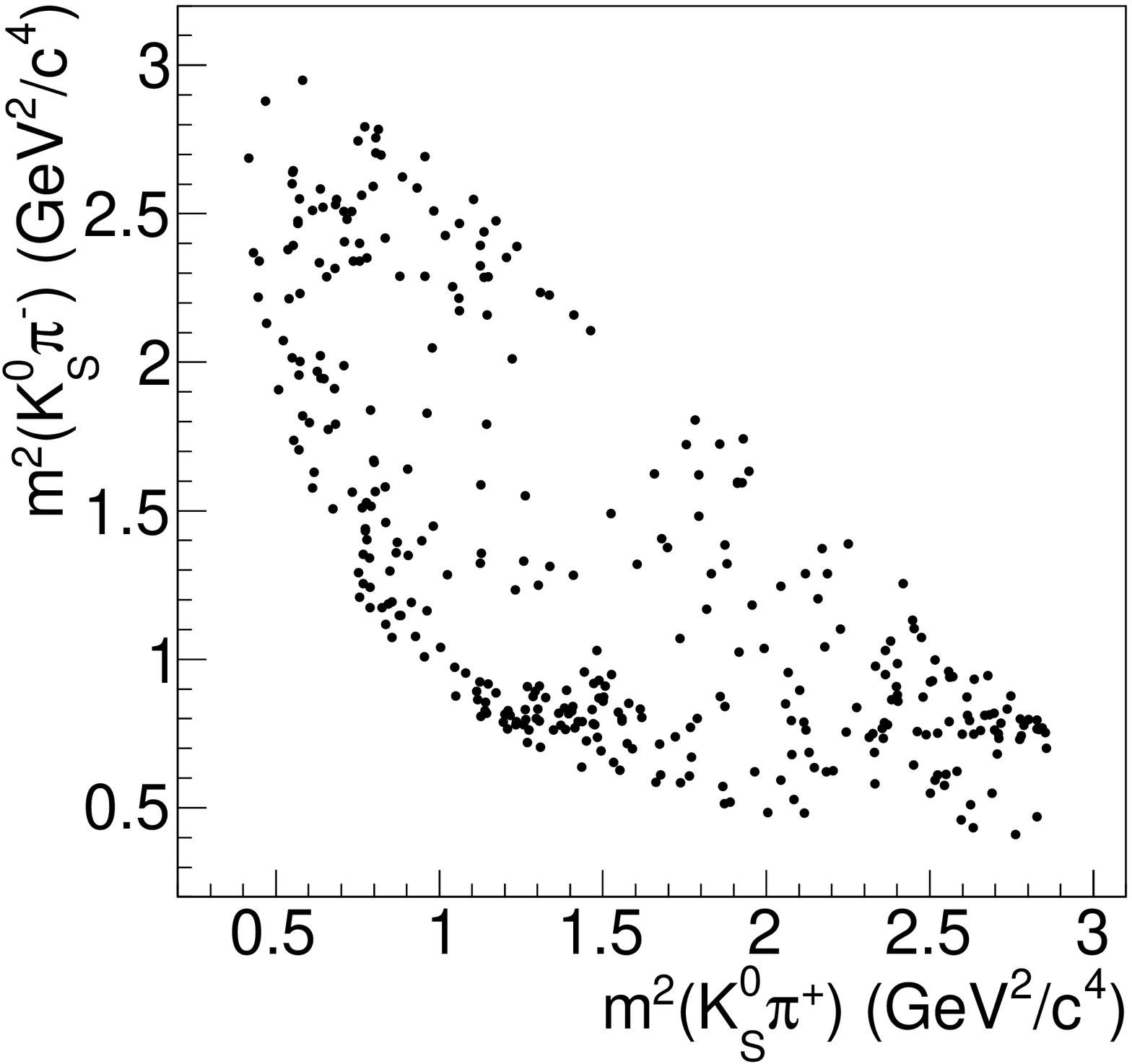}}
\caption{Dalitz distributions for $D$ mesons produced in {tagged} \bdsth decays with 
wrong tag probability of less than $23\%$. The signal $B$ meson is tagged as \bn~(a) and \bnbar~(b).}
\label{fig:dp-bdh}
\end{figure*}

The ellipses in the \dembc plane inscribed in the rectangular areas marked by the 
vertical red lines in Figs.~\ref{fig:de-mbc}~and~\ref{fig:de-mbc_other} are defined 
for each signal mode and are referred to as signal regions.  The events in these signal 
regions are used in the fit of the \cpconj violation parameters.  The signal yields 
$N\subsig$ and fractions $f\subsig$ of signal events for each signal region obtained 
from the \dembc fit, are listed in~Table~\ref{tab:data_de_mbc}.  The Dalitz plots for 
events with a wrong-tag probability of under $23\%$ are shown in~Fig.~\ref{fig:dp-bdh}.

\section{\boldmath Determination of the \cpconj violation parameters}
\begin{figure}[htb]
\subfloat[]{\label{fig:acp1}\includegraphics[width=0.23\textwidth]{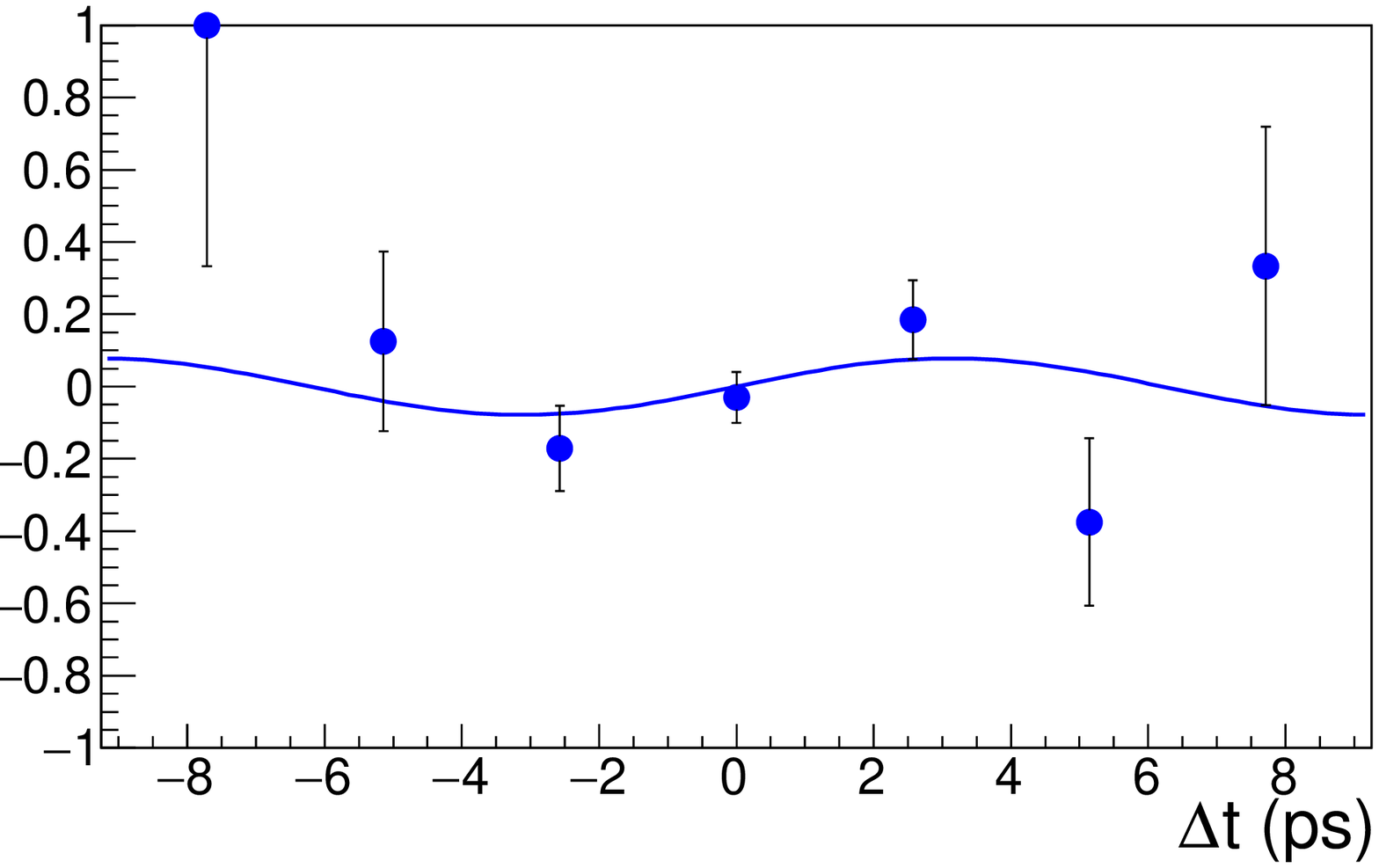}}\hfill
\subfloat[]{\label{fig:acp5}\includegraphics[width=0.23\textwidth]{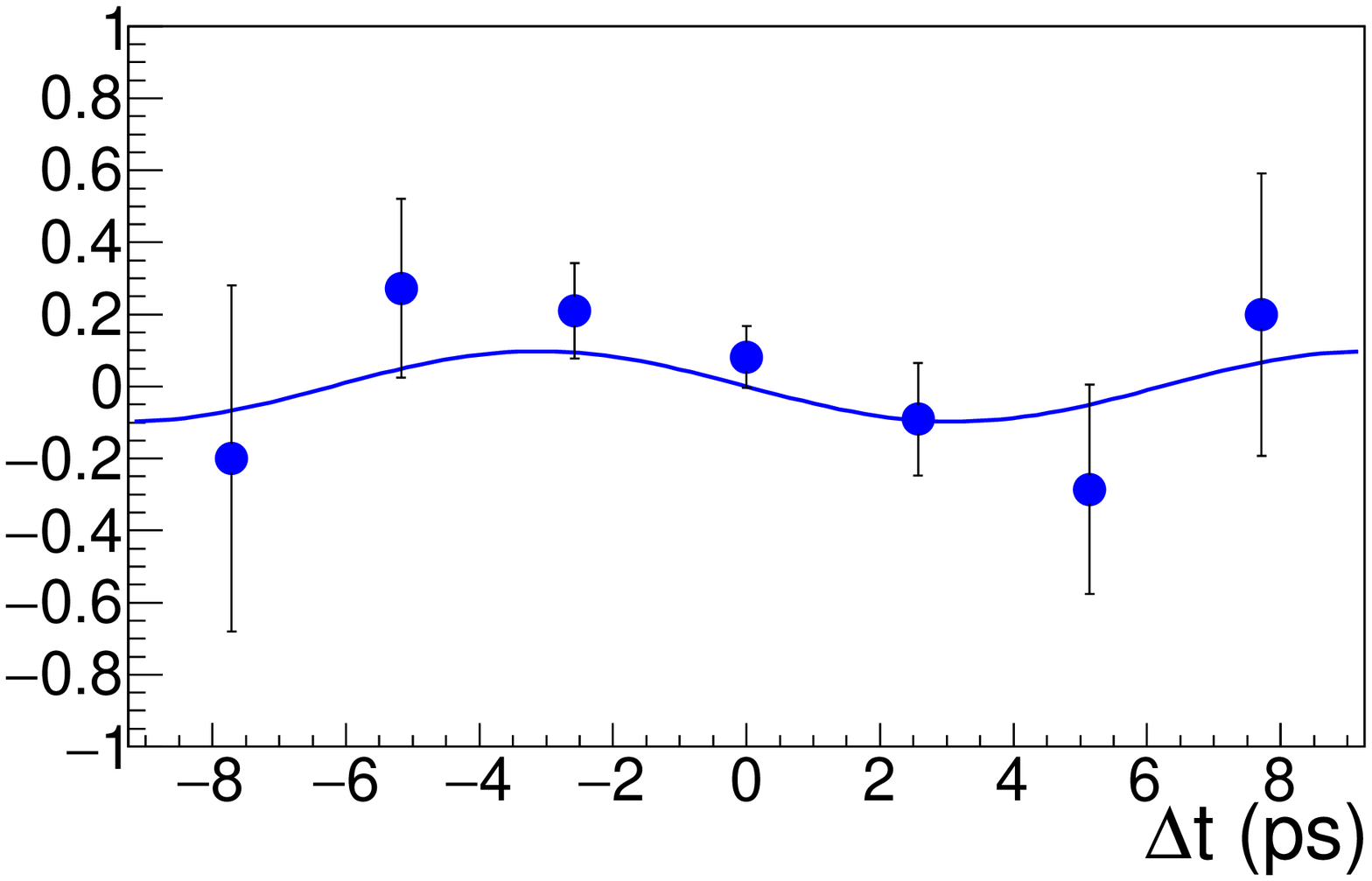}}
\caption{Raw \cpconj asymmetry distributions for the \bdsth candidates in the $\pm1$-st 
(a) and $\pm5$-th (b) \dkpp decay Dalitz plot bins.  Red lines are the result of the 
\cpconj violation fit performed with the full data sample.  The asymmetry for the 
\bdstarh candidates is taken with inverted sign.}
\label{fig:acp}
\end{figure}

\begin{figure}[ht]
\subfloat[]{\label{fig:dtdist3}\includegraphics[width=0.23\textwidth]{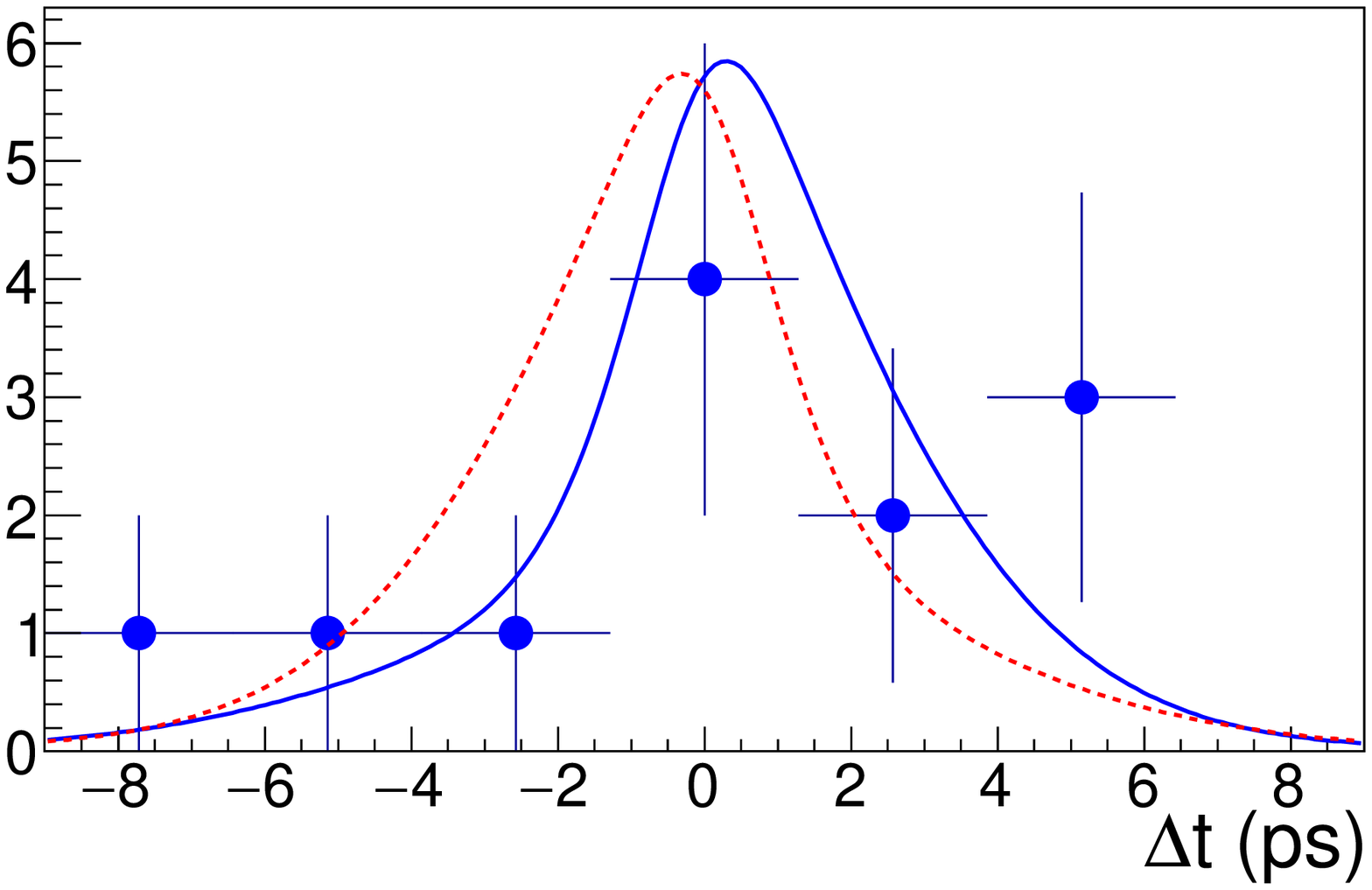}}\hfill
\subfloat[]{\label{fig:dtdist7}\includegraphics[width=0.23\textwidth]{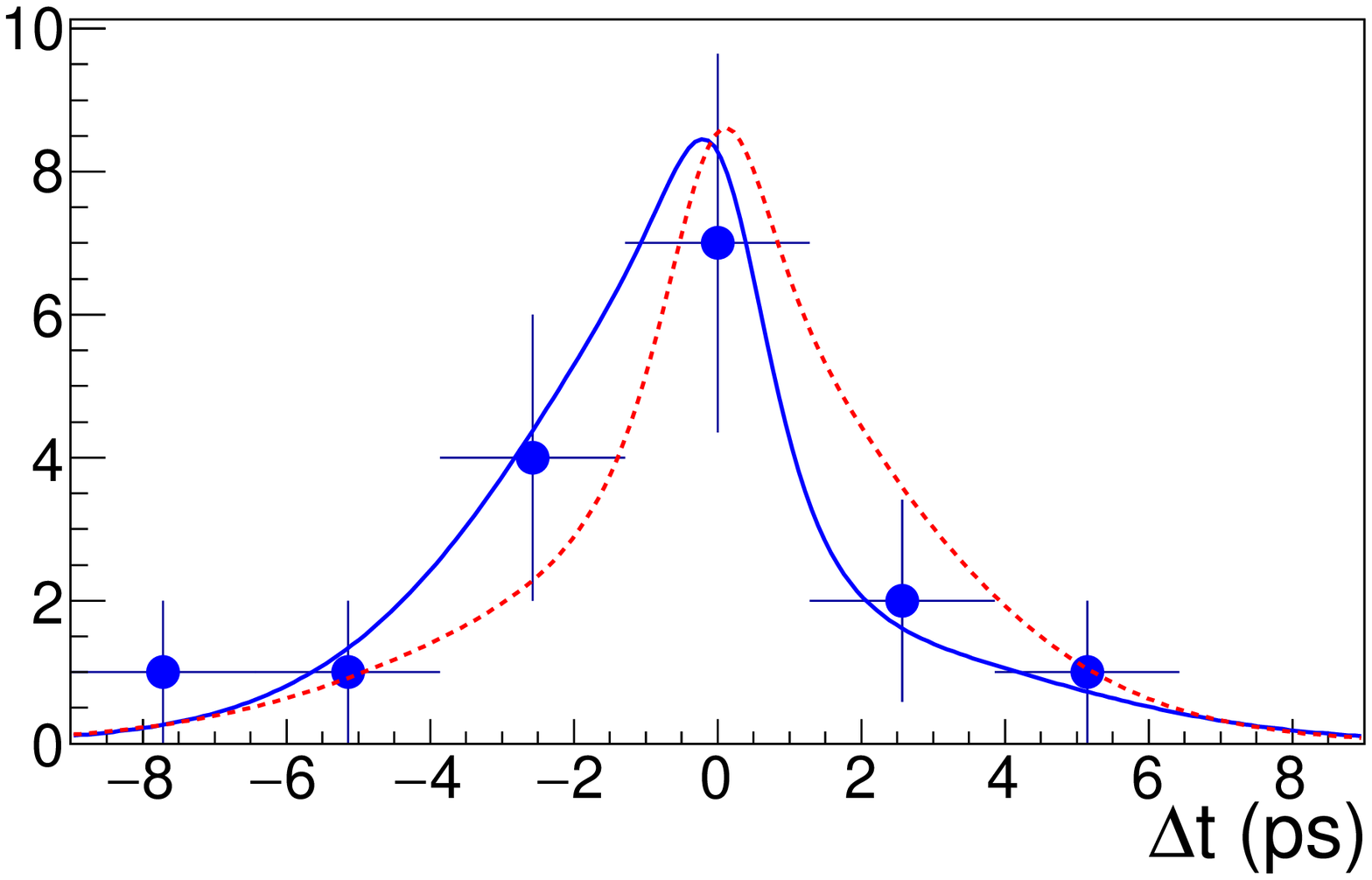}}
\caption{\dt distributions for the \bdh candidates with wrong-tag probability of less than $23\%$. 
(a)~((b)) corresponds to the candidates from the $-3$-th ($7$-th) \dkpp Dalitz plot bin tagged as 
\bn (\bnbar).  Continuous blue lines are the result of the \cpconj violation fit performed with 
the full data sample.  Dashed red lines are obtained with $\pphi=68.1\grad$.}
\label{fig:dtdist}
\end{figure}

The \cpconj violation parameters are measured using the unbinned maximum 
likelihood fit of the \dt distribution. The likelihood function is defined as
\begin{equation}\label{eq:lh}
 \mcl = \prod\limits_{j=1}^{N}\left[(1-f_{{\rm bkg},j})p\subsig(\dt_j)+f_{{\rm bkg},j}p\subbkg(\dt_j)\right],
\end{equation}
where the product is evaluated over all $N$ events in the sample, $f_{{\rm bkg},j}$ 
is the event-dependent background fraction obtained from the \dembc fit, 
$p\subsig$ is the signal PDF, and $p\subbkg$ is the background PDF.

The background \dt distributions are parameterized by convolving the function
\begin{equation}
 f_{\delta}\delta(\dt)+\left(1-f_{\delta}\right)2\tau\subbkg e^{-|\dt|/\tau\subbkg}
\end{equation}
with a double-Gaussian function; here, $\delta$ is the Dirac delta function and 
$\tau\subbkg$ is the effective lifetime for background events.  The widths of the 
double-Gaussian function are event-dependent and proportional to the estimated 
vertex resolution obtained from the vertex-constrained kinematic fits.  The 
parameters $f_{\delta}$ and $\tau\subbkg$ are obtained from simulation while 
the parameters of the double-Gaussian function are obtained from the fit of the 
\dt distribution in the \dembc sideband. The \dt distributions for background 
from \bbbar events and from continuum events are parameterized separately.  

The signal \dt distribution is parameterized by convolving Eq.~(\ref{eq:master-formula}) 
with a resolution function. The resolution function is described in~Ref.~\cite{vertexres}.  
It is tuned for each event using information obtained from the vertex-constrained kinematic fits.  

Table~\ref{tab:data_cpv} shows results of the fit of the \cpconj violation parameters, 
where \sindbeta and \cosdbeta are treated as independent variables.  The 
correlation coefficient of \sindbeta and \cosdbeta is about $-3\%$.  

\begin{table}[htb]
\caption{Fit of the \cpconj violation parameters. Only statistical uncertainty is shown.}
\label{tab:data_cpv}
\begin{tabular}
{ @{\hspace{0.5cm}}l@{\hspace{0.5cm}} @{\hspace{0.5cm}}c@{\hspace{0.5cm}} @{\hspace{0.5cm}}c@{\hspace{0.5cm}} } \hline\hline
Signal mode   & $\sindbeta$              & $\cosdbeta$            \\ \hline 
\bdpi         & $\phantom{-}0.61\pm0.37$ & $0.88^{+0.46}_{-0.52}$ \\
\bdomega      & $         - 0.12\pm0.58$ & $1.28^{+0.62}_{-0.69}$ \\ 
Other modes   & $\phantom{-}0.44\pm0.51$ & $0.89^{+0.49}_{-0.55}$ \\ \hline
All modes     & $\phantom{-}0.41\pm0.27$ & $0.97\pm0.33$          \\ \hline \hline
\end{tabular}
\end{table}

The combined fit of all signal modes, with the parameters \sindbeta and \cosdbeta 
considered as functions of the angle \pphi, results in
\begin{equation}\label{eq:phi1_stat}
 \pphi = 11.7\grad\pm7.8\grad\stat.
\end{equation}

For illustration, the raw \cpconj asymmetries for the Dalitz plot bins most sensitive to \sindbeta 
are shown in Fig.~\ref{fig:acp}.  The \dt distributions for the Dalitz plot bins most sensitive 
to \cosdbeta are shown in Fig.~\ref{fig:dtdist}.

\section{Systematic uncertainties}
\begin{figure*}[htb]
\subfloat[]{\label{fig:sin_minos}\includegraphics[width=0.32\textwidth]{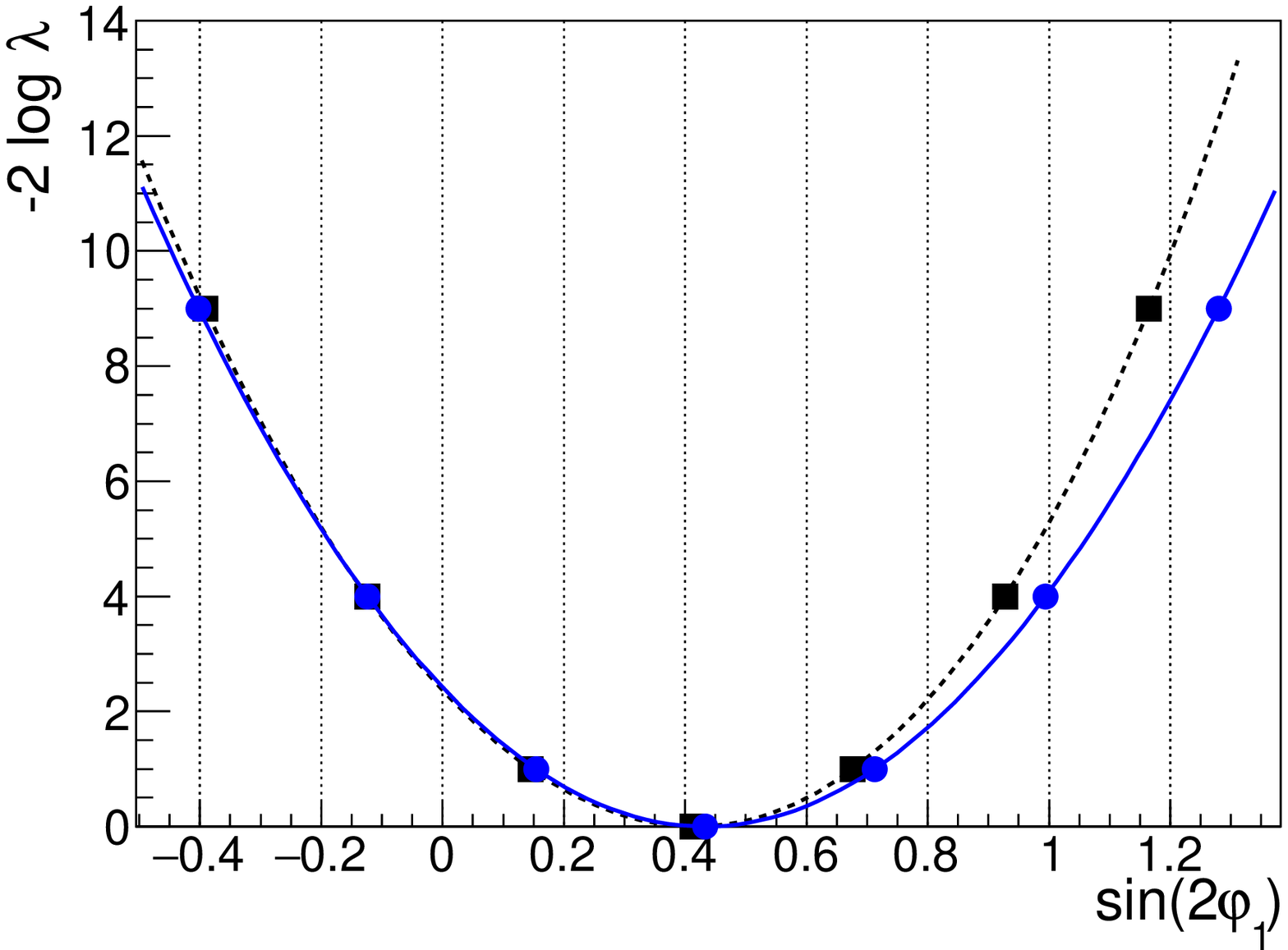}}\hfill
\subfloat[]{\label{fig:cos_minos}\includegraphics[width=0.32\textwidth]{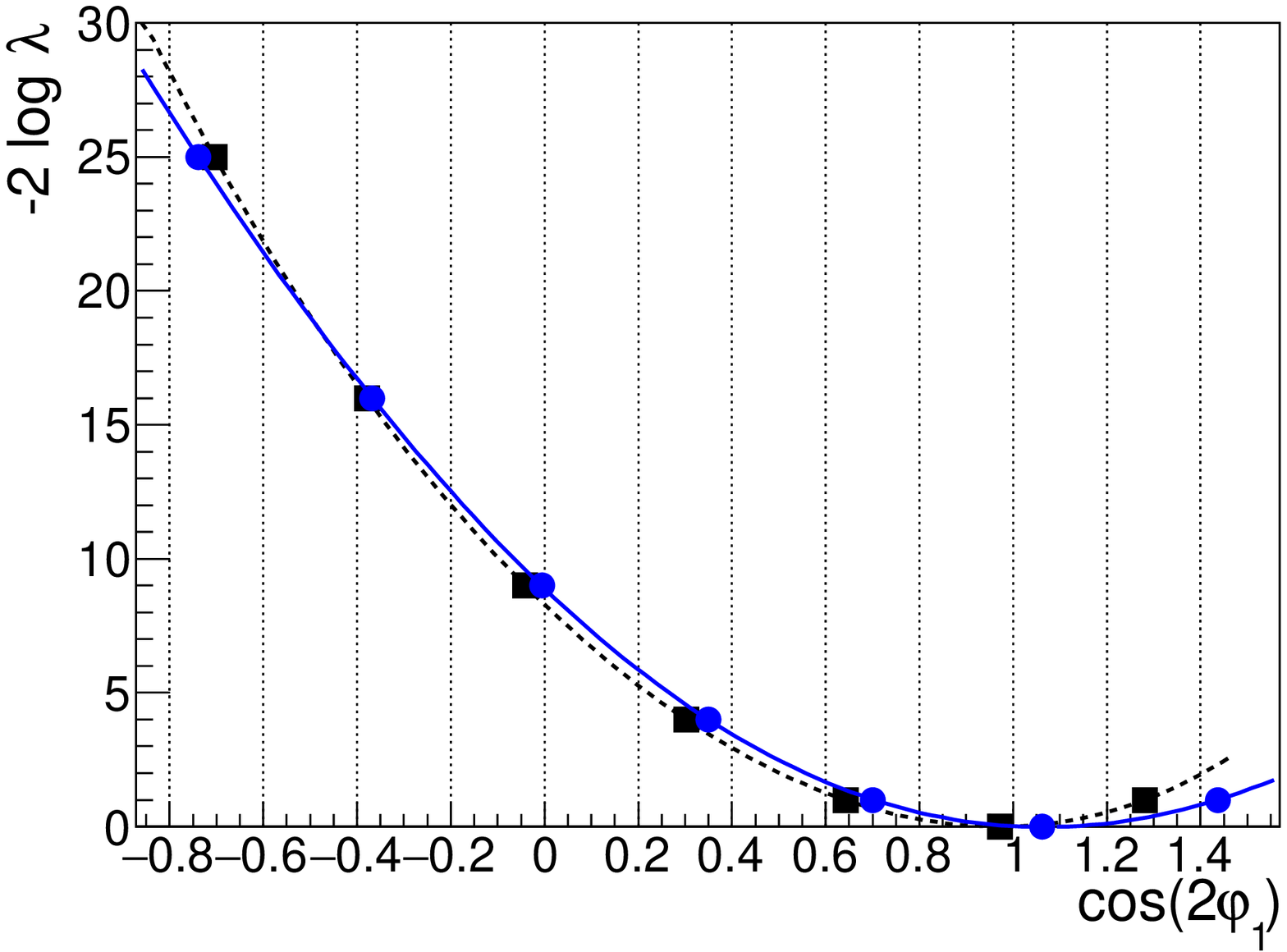}}\hfill
\subfloat[]{\label{fig:phi1minos}\includegraphics[width=0.32\textwidth]{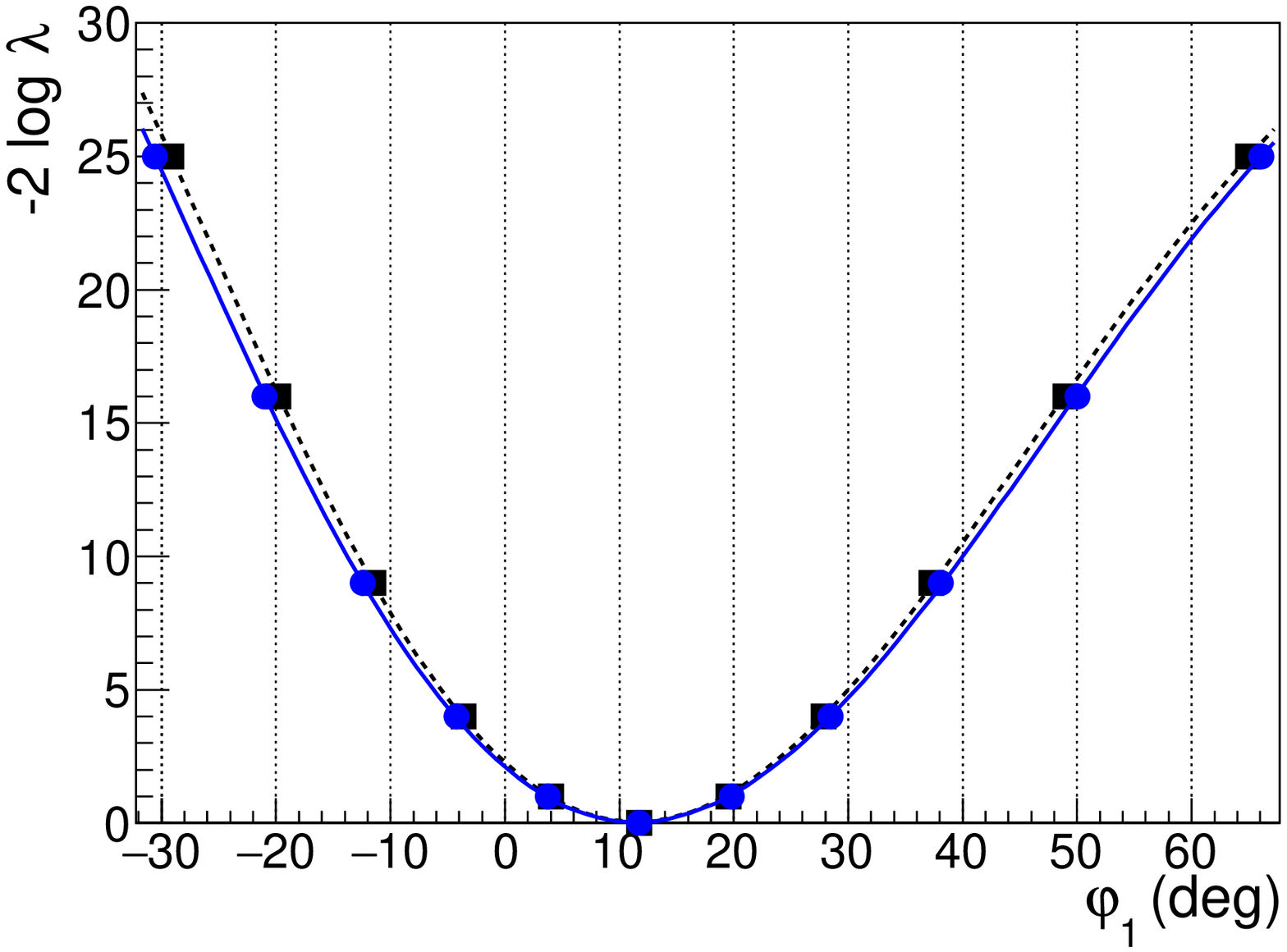}}
\caption{Negative double logarithm of profiled likelihood ratio 
Eq.~(\ref{eq:confidence_intervals}) for \sindbeta~(a), \cosdbeta~(b) and 
\pphi~(c) obtained with the Minos algorithm~\cite{minos}. Black squares mark 
$n\sigma$ standard confidence intervals corresponding to statistical 
uncertainty while blue circles mark $n\sigma$ standard confidence 
intervals corresponding to the overall uncertainty. Continuous blue and 
dashed black lines show $4$-th (a), (b) and  $5$-th~order (c)
polynomial fit.}
\label{fig:cpv_minos}
\end{figure*}

Table~\ref{tab:systematics_cpv} provides the estimates of the systematic 
uncertainties in the measured values of the \cpconj violation parameters.

The uncertainty due to the experimental resolution for the Dalitz variables 
is evaluated using the large sample of simulated signal events.  The fit 
results are compared for the the \cpconj violation fit performed using 
the reconstructed and the generated Dalitz-variables values.  The uncertainty 
due to the detection-efficiency variation over the Dalitz plot is also evaluated 
using the simulated signal events.  The fit results are compared for the 
\cpconj violation fit performed with and without the efficiency correction.  

The systematic uncertainty related to the signal \dt parameterization is estimated 
by varying each resolution parameter by $\pm\sigma_{\pm}$ ($\pm2\sigma_{\pm}$ 
for parameters obtained from MC simulation) and repeating the fit.  

Other contributions to the systematic uncertainty (items $4$\,--\,$10$ in 
Table~\ref{tab:systematics_cpv}) are evaluated simultaneously from the fit performed 
with \emph{nuisance} parameters and the likelihood function expressed as follows:
\begin{equation}\label{eq:nuislh}
 -2\log{\mcl_{\rm n}} = -2\log{\mcl}+\sum\limits_{j,k}\left(p_j-p_j^{0}\right)\mck_{jk}\left(p_k-p_k^{0}\right),
\end{equation}
where \mcl is defined in Eq.~(\ref{eq:lh}), $p_j$ and $p_j^0$ are the 
current and central values of the $j$-th nuisance parameter, respectively, 
\mck is the inverse covariance matrix for the nuisance parameters 
and the sum is evaluated over all nuisance parameters.  The following 
nuisance parameters are introduced to evaluate the systematic uncertainty:
\begin{itemize}
 \item the parameters \ci and \si that give the dominant contribution (with 
 the covariance matrix taken from the supplementary  materials for 
 Ref.~\cite{CLEO_phasees}).
 \item the parameters \ki with the uncertainties shown in Table~\ref{tab:Kmeasured};
 \item the yield of signal events in each Dalitz plot bin for each signal 
 mode with the value and uncertainty obtained from the \dembc fit;
 \item the background \dt PDF parameters with the values and uncertainties 
 obtained from the fit of the \dt distribution in the \dembc sideband;
 \item the parameters \btau and \dmb with values and uncertainties taken from~Ref.~\cite{PDG}; and
 \item the average bias in the wrong-tag probability with the uncertainty obtained using the 
 results from~Ref.~\cite{Tagging}.
\end{itemize}

The flavor tagging procedure and the uncertainties in the \dmb and \btau values 
give negligible contributions to the systematic uncertainty.  

Frequentist confidence intervals for the \cpconj violation parameters are evaluated 
using the profile likelihood method with likelihood ratios~\cite{statistics}
\begin{equation}\label{eq:confidence_intervals}
 \lambda(\xi)=\frac{\mcl(\xi,\hat{\hat{p}})}{\mcl(\hat{\xi},\hat{p})},
\end{equation}
where $\xi$ is \sindbeta or \cosdbeta or \pphi, $\hat{\xi}$ is the optimal 
value, $\hat{p}$ represents the optimal values of all other parameters corresponding 
to $\hat{\xi}$, and $\hat{\hat{p}}$ represents the optimal values of all other 
parameters corresponding to the $\xi$ value.  Negative double logarithms of the 
likelihood ratios are shown in~Fig.~\ref{fig:cpv_minos}.  

\begin{table}[htb]
\caption{The sources and estimates of the systematic uncertainties for the \cpconj 
violation parameters measured in the \bdsth decays.  The uncertainty $\sigma_{\rm nuis}$ 
due to the sources $4$\,--\,$10$ is evaluated from the single fit varying all the 
nuisance parameters and using the likelihood function~Eq.~(\ref{eq:nuislh}).  The total 
systematic uncertainty $\sigma_{\rm syst}$ is calculated as $\sqrt{\sigma_1^2+
\sigma_2^2+\sigma_3^2+\sigma_{\rm nuis}^2}$.  The values related to the sources 
$4$\,--\,$10$ are shown for illustration.}
\label{tab:systematics_cpv}
\begin{tabular}
{ @{\hspace{0.cm}}l@{\hspace{0.cm}} @{\hspace{0.cm}}c@{\hspace{0.1cm}} @{\hspace{0.1cm}}c@{\hspace{0.1cm}} @{\hspace{0.1cm}}c@{\hspace{0.cm}} } \hline\hline
 Source           & $\delta_{\sindbeta}$\,($\%$) & $\delta_{\cosdbeta}$\,($\%$) & $\delta_{\pphi}$\,(deg)\\ \hline 
1. Dalitz variables resol.& $0.3$           & $0.7$             & $0.1$ \\
2. Detection efficiency   & $0.6$           & $0.8$             & $0.2$ \\
3. \dt resolution        & $3.8$           & $6.7$             & $1.2$ \\ 
4. Flavor tagging         & $0.1$           & $0.1$             & $<0.1$ \\
5. \dmb                   & $0.1$           & $0.1$             & $<0.1$ \\
6. \btau                  & $0.1$           & $0.1$             & $<0.1$ \\
7. \mbc-\de fit           & $3.4$           & $1.9$             & $0.8$ \\
8. Bkg. \dt param.        & $3.6$           & $3.1$             & $0.7$ \\
9. \ki                    & $3.2$           & $2.0$             & $0.7$ \\
10. \ci and \si           & $7.6$           & ${}^{+20}_{-13}$  & $1.1$ \\ \hline
$\sigma_{\rm nuis}$       & $7.6$           & ${}^{+20}_{-13}$  & $1.6$ \\
Total $\sigma_{\rm syst}$ & $8.5$           & ${}^{+21}_{-15}$  & $2.1$ \\
{\it Stat. error for comparison}  & $\mathit{27}$ & $\mathit{33}$ & $\mathit{7.8}$ \\ \hline
\hline
 \end{tabular}
\end{table}

The dominant uncertainties shown in Table~\ref{tab:systematics_cpv} could be reduced in 
high-statistics measurements at Belle\,II.  Indeed, the uncertainties associated with the 
parameters \ki, the \dt parameterization and the \dembc fit are determined by the size of 
the data sample.  The parameters \ci and \si can be measured more precisely with a large 
data set of coherently produced \ddbar pairs collected by the BES\,-\,III experiment.

\section{Conclusions}
A novel model-independent approach for measuring the CKM angle \pphi has been 
developed and applied to the full data set of the Belle experiment. The following 
results are obtained:
 \begin{equation}
 \begin{split}
  \sindbeta &= 0.43     \pm 0.27\stat    \pm 0.08    \syst,\\
  \cosdbeta &= 1.06     \pm 0.33\stat^{+0.21}_{-0.15}\syst,\\
  \pphi     &= 11.7\grad\pm 7.8\grad\stat\pm 2.1\grad\syst.
 \end{split}
 \end{equation}

The value $\sindbeta~=~0.691~\pm~0.017$ measured in \btoccs transitions determines 
the absolute value of \cosdbeta leading to two possible solutions in the 
$0\grad\leq\pphi<180\grad$ range.  Our measurement is inconsistent with the 
negative solution corresponding to the value $\pphi = 68.1\grad$ at the level 
of $5.1$ standard deviations but in agreement with the positive solution corresponding 
to the value $\pphi = 21.9\grad$ at $1.3$ standard deviations.  Thus, this measurement 
clearly resolves the ambiguity in \pphi inherent in the measurement of \sindbeta using 
the \btoccs transition.

This measurement supersedes the previous measurement of the \sindbeta and \cosdbeta in 
\bdsth decays at Belle~\cite{belle_bdh0}.  Nevertheless, it should be emphasized that a different 
analysis technique is used here.  Furthermore, experimental information from \bptodpi decays 
and from Ref.~\cite{CLEO_phasees} is used in this analysis but not in Ref.~\cite{belle_bdh0}.

The binned Dalitz plot approach could be used for precise \pphi measurements in \bdsth 
followed by \dbkpp decays with the high-statistics data from the Belle~II experiment.  
The dominant systematic uncertainties could be reduced with this larger data sample.  
Also, abundant coherently-produced \ddbar pairs collected by the BES\,-\,III experiment can 
be used to improve our knowledge of the phase parameters \ci and \si. The number of 
Dalitz plot bins can be increased in future measurements to improve the statistical 
sensitivity to the \cpconj violation parameters.  

Some NP models predict the magnitude of \cpconj violation to differ from the SM 
expectations~\cite{acp_np}.  The difference may vary for different quark transitions.  
Thus, it would be interesting to compare the \sindbeta value precisely measured in the \btocud 
transitions governing the \bdsth decays with the \sindbeta value precisely measured in the 
\btoccs transitions.

\section{Acknowledgments}

We thank the KEKB group for the excellent operation of the
accelerator; the KEK cryogenics group for the efficient
operation of the solenoid; and the KEK computer group,
the National Institute of Informatics, and the 
PNNL/EMSL computing group for valuable computing
and SINET4 network support.  We acknowledge support from
the Ministry of Education, Culture, Sports, Science, and
Technology (MEXT) of Japan, the Japan Society for the 
Promotion of Science (JSPS), and the Tau-Lepton Physics 
Research Center of Nagoya University; 
the Australian Research Council;
Austrian Science Fund under Grant No.~P 22742-N16 and P 26794-N20;
the National Natural Science Foundation of China under Contracts 
No.~10575109, No.~10775142, No.~10875115, No.~11175187, No.~11475187
and No.~11575017;
the Chinese Academy of Science Center for Excellence in Particle Physics; 
the Ministry of Education, Youth and Sports of the Czech
Republic under Contract No.~LG14034;
the Carl Zeiss Foundation, the Deutsche Forschungsgemeinschaft, the
Excellence Cluster Universe, and the VolkswagenStiftung;
the Department of Science and Technology of India; 
the Istituto Nazionale di Fisica Nucleare of Italy; 
the WCU program of the Ministry of Education, National Research Foundation (NRF) 
of Korea Grants No.~2011-0029457,  No.~2012-0008143,  
No.~2012R1A1A2008330, No.~2013R1A1A3007772, No.~2014R1A2A2A01005286, 
No.~2014R1A2A2A01002734, No.~2015R1A2A2A01003280 , No. 2015H1A2A1033649;
the Basic Research Lab program under NRF Grant No.~KRF-2011-0020333,
Center for Korean J-PARC Users, No.~NRF-2013K1A3A7A06056592; 
the Brain Korea 21-Plus program and Radiation Science Research Institute;
the Polish Ministry of Science and Higher Education and 
the National Science Center;
the Ministry of Education and Science of the Russian Federation and
the Russian Foundation for Basic Research;
the Slovenian Research Agency;
Ikerbasque, Basque Foundation for Science and
the Euskal Herriko Unibertsitatea (UPV/EHU) under program UFI 11/55 (Spain);
the Swiss National Science Foundation; 
the Ministry of Education and the Ministry of Science and Technology of Taiwan;
and the U.S.\ Department of Energy and the National Science Foundation.
This work is supported by a Grant-in-Aid from MEXT for 
Science Research in a Priority Area (``New Development of 
Flavor Physics'') and from JSPS for Creative Scientific 
Research (``Evolution of Tau-lepton Physics'').

\end{document}